\documentclass[floatfix,aps,prb,twocolumn,superscriptaddress,
10pt]{revtex4-2}

\usepackage{graphicx}
\usepackage{ulem}
\usepackage{color}
\usepackage{physics}
\usepackage{booktabs}
\usepackage{array}
\usepackage{listings}
\usepackage{soul}
\usepackage{float}
\usepackage[T1]{fontenc}
\usepackage[utf8]{inputenc}
\usepackage{caption}
\usepackage{subcaption}
\usepackage{afterpage}
\usepackage{xcolor}
\usepackage{hyperref}
\hypersetup{colorlinks=true, linkcolor=blue, citecolor=blue, urlcolor=blue}

\begin{document}

\widetext
\title{First-principles study of lattice dynamical properties of the room-temperature $P2_1/n$ and ground-state $P2_1/c$ phases of WO$_3$ }

\author{Hamideh Hassani}
\affiliation{Physique Th\'eorique des Mat\'eriaux, QMAT, CESAM, Universit\'e de L\`iege, B-4000 Sart-Tilman, Belgium}
\affiliation{Department of Physics, University of Antwerp, Groenenborgerlaan 171, 2020 Antwerp, Belgium}

\author{Bart Partoens}
\affiliation{Department of Physics, University of Antwerp, Groenenborgerlaan 171, 2020 Antwerp, Belgium}

\author{Eric Bousquet}
\affiliation{Physique Th\'eorique des Mat\'eriaux, QMAT, CESAM, Universit\'e de L\`iege, B-4000 Sart-Tilman, Belgium}

\author{Philippe Ghosez}
\affiliation{Physique Th\'eorique des Mat\'eriaux, QMAT, CESAM, Universit\'e de L\`iege, B-4000 Sart-Tilman, Belgium}

\date{\today}

\begin{abstract}

Using first-principles density functional theory, we investigate the dynamical properties of the room-temperature $P2_1/n$ and ground-state $P2_1/c$ phases of WO$_3$.  As a preliminary step, we assess the validity of various standard and hybrid functionals, concluding that the best description is achieved with the B1-WC hybrid functional  while a reliable description can also be provided using the standard LDA functional. We also carefully rediscuss the structure and energetics of all experimentally observed and few hypothetical metastable phases in order to provide deeper insight into the unusual phase diagram of WO$_3$. Then, we provide a comprehensive theoretical study of the lattice dynamical properties of the $P2_1/n$ and $P2_1/c$ phases, reporting zone-center phonons, infrared and Raman spectra as well as the full phonon dispersion curves, which attest for the dynamical stability of both phases. We carefully discuss the spectra, explaining the physical origin of their main features and evolution from one phase to another. We reveal a systematic connection between the dynamical  and structural properties of WO$_3$, highlighting that the number of peaks in the high-frequency range of the Raman spectrum appears as a fingerprint of the number of antipolar distortions that are present in the structure and a practical way to discriminate between the different phases. 
\end{abstract}
\maketitle

\section{Introduction}
Tungsten trioxide, WO$_3$, is a wide bandgap semiconductor transition metal oxide with peculiar electronic, chromic and optical properties that makes it appealing for photocatalytic, electrochromic and optoelectrical devices applications \cite{Niklasson2007,Deb2008}. WO$_3$ can be classified as a perovskite-like ABO$_3$ compound but with an empty $A$ site. As such, it possesses an aristotype $Pm\bar{3}m$ cubic structure, in which W atoms sit at the center of oxygen octahedra sharing a corner and forming a continuous network. However, WO$_3$ is typically not observed in that reference cubic structure and experimental measurements at various temperatures reveal a very complex phase diagram, not fully elucidated yet \cite{Howard2002}. At about 1300 K, WO$_3$ is reported to crystallize in a $P4/nmm$ tetragonal phase ~\cite{Howard2002,Locherer_1999,Kehl1952}. Then, on cooling, it shows consecutive phase transitions to (i) a $P4/ncc$ tetragonal phase at about 1150 K~\cite{Howard2002,Locherer_1999,Vogt1999,Kehl1952}, (ii) an eventual $P2_1/c$ monoclinic phase at about 1050 K ~\cite{Howard2002}, (ii) a $Pbcn$ orthorhombic phase at about 1000 K~\cite{Vogt1999,Howard2002,Locherer_1999,Salje:1977}, (iii) a $P2_1/n$ monoclinic phase at about 600 K~\cite{Howard2002,Locherer_1999,Vogt1999,Loopstra:1969}, (iv) a $P\overline{1}$ triclinic phase at 273 K~\cite{Diehl:a15962,Locherer_1999,Woodward1995} and, ultimately, (v) a $Pc$ monoclinic polar phase at around 200 K~\cite{Salje1997,Locherer_1999,WOODWARD19979,salje1976} that was recently proposed to be instead a related non-polar $P2_1/c$ monoclinic phase \cite{Hamdi2016}. All these phases appear as small distortions of the cubic artistotype structure but including distinct rotations and tilts of the oxygen octahedra and different shifts of W atoms within these octahedra. 

Experimentally, WO$_3$ appears also in practice as a sub-stoichiometric (WO$_{3-x}$) compound. This, on the one hand, confers it some unique properties exploited in a variety of technological applications \cite{Migass2010} but, on the other hand, makes its study significantly more complex. In fact, the off-stoichiometry makes it a $n$-doped semiconductor and the excess electrons interact with the polarizable crystal lattice in order to form polarons and bipolarons \cite{Bousquet2020} that strongly influence its properties. 

Although extensive studies have been reported concerning the electronic and optical properties of WO$_3$ \cite{Johansson2013,Gonzalez-Borrero2010,Ping2013,Migass2010}, there is a much more limited amount of works dedicated to the investigation of its lattice dynamical properties. The phonon dispersion curves of cubic WO$_3$  and its elastic properties have been studied by Fan {\it et al.} \cite{Fan2018} or together with the electron-phonon coupling by Mascello et al.~\cite{mascello2020}. 
Some lattice dynamical properties of distinct WO$_3$ phases have also been reported by Yang {\it et al.} ~\cite{Yang2019}. Furthermore, Hamdi \textit{et al.}~\cite{Hamdi2016} studied the phonon instabilities of the cubic phase together with the various potential metastable phases originating from these instabilities. This latter study has shown that the B1 Wu-Cohen (B1-WC) hybrid exchange-correlation functional~\cite{Bilc2008} can accurately reproduce the experimental results regarding both the electronic and structural properties of WO$_3$. 

In the present work, we first carefully re-discuss the phase diagram of WO$_3$, assessing the validity of distinct functionals. We then provide a comprehensive theoretical study of the lattice dynamical properties of the $P2_1/n$ room temperature phase and $P2_1/c$ expected ground-state phase, reporting infrared and Raman spectra together with the full phonon dispersion curves. On the one hand, we carefully discuss the spectra, explaining the physical origin of their main features and evolution from one phase to another and providing meaningful benchmark results for the interpretation of experimental measurements. On the other hand, polaron formation is directly linked to electron-phonon interaction so that the present understanding of the phonon properties appears as a useful step toward the understanding of polarons in these two phases.

The paper is organized as follows. In the first part, we compare the structural parameters and energetics of various phases of WO$_3$, as obtained using standard (LDA and GGA) and more advanced (B1-WC, HSE06) hybrid functionals. We also take this opportunity to re-discuss briefly the phase diagram of WO$_3$.  In the second part, we carefully discuss the lattice dynamical properties of the $P2_1/n$ room temperature and  $P2_1/c$ ground state structures, comparing our theoretical results with experimental data when available and interpreting the shape of the spectra. From this analysis, Raman measurements emerges as a promising tool to identify the various phases of WO$_3$.

\section{Computational details}

Our calculations are performed in the framework of density functional theory (DFT) as implemented in CRYSTAL17~\cite{Caus1987} based on a linear combination of localized basis functions and ABINIT~\cite{Gonze2009,Gonze2002} making use a plane-wave basis set. Calculations are reported using distinct approximations of the exchange-correlation energy including the local density approximation (LDA), the generalized gradient approximation (GGA) and different hybrid schemes. Integrations over the Brillouin zone are approximated by sums over a mesh of $8 \times 8 \times 8 $ $k$-points for the cubic phase or meshes providing a comparable sampling for other phases (meshes of $6 \times 6 \times 8$, $6 \times 6 \times 4$ and $4 \times 4 \times 4$ are used for \textit{P4/nmm}, \textit{P4/ncc} and \textit{Pbcn} phases respectively).

ABINIT is used with the LDA with Perdew-Wang’s parametrization \cite{PhysRevB.45.13244} and the GGA with both the PBEsol \cite{Perdew2008} and Wu and Cohen (WC) functionals \cite{Wu2006}. These calculations make use of norm–conserving pseudopotentials from the Pseudo Dojo Table v0.4~\cite{Setten2018}, considering the $2s$ and $2p$ orbitals of O and the $5s$, $5p$, $5d$ and $6s$ orbitals of W as valence states. The electronic wave functions are expanded in plane-waves up to an energy cutoff of 60 Ha. Full relaxations of the lattice parameters and atomic positions are performed  with each functional using the Broyden-Fletcher-Goldfarb-Shanno (BFGS) algorithm \cite{10.1093} until the maximal forces and stresses are less than $10^{-6}$ Ha/Bohr and $10^{-4}$ GPa, respectively. The electronic self-consistent calculations were converged until the difference of the total energy is smaller than $10^{-9}$ Ha. Dynamical matrices, Born effective charges and dielectric tensors are obtained within a variational approach \cite{PhysRevLett.68.3603} to density functional perturbation theory (DFPT) \cite{PhysRevLett.58.1861}.

CRYSTAL is used for the hybrid functional calculations with both the B1-WC functional \cite{PhysRevB.77.165107} and the HSE06 functional \cite{doi:10.1063/1.1564060}. Benchmark calculations in LDA have also been performed for comparison with ABINIT. We use all-electron double-$\zeta$ basis sets for oxygen whereas, for the heavy tungsten atom, we use the effective core pseudopotential (ECP) technique as implemented in the small core Detlev-Figgen pseudopotentials associated with correlation consistent polarized Valence Triple-$\zeta$ Basis Set (cc-pVDZ) \cite{Figgen2009} for explicit treatment of valence electrons. The truncation thresholds in the evaluation of Coulomb and exchange series appearing in the SCF equation for periodic systems are adjusted to ${10}^{-7}$ for Coulomb overlap tolerance, ${10}^{-7}$ for Coulomb penetration tolerance, ${10}^{-7}$ for exchange overlap tolerance, ${10}^{-7}$ for exchange pseudo-overlap in the direct space, and ${10}^{-14}$ for exchange pseudo-overlap in the reciprocal space. The tolerance on change in total energy for the SCF convergence is adjusted to ${10}^{-10}$ Ha. Full structural relaxations are carried out by using a quasi-Newton algorithm with a BFGS Hessian updating scheme so that a convergence of less than $5 \times 10^{-5}$ Ha/Bohr and $10^{-3}$ Bohr is reached respectively on the root mean square of the gradient and displacements.  The phonon frequencies are computed using the frozen phonon numerical differences \cite{jcc.20019, jcc.20120} (for detailed information check Fig.~\ref{fig:A1a} in the appendix). The Born effective charges and the optical dielectric tensors are evaluated via the Berry phase technique \cite{PhysRevB.47.1651} and the Coupled-Perturbed Kohn-Sham approach \cite{1.3267861, 1.3043366}, respectively.

\noindent
\section{Structural properties}

\begin{figure}[tb]
\hspace*{-1cm} 
\includegraphics[width=0.5\textwidth]{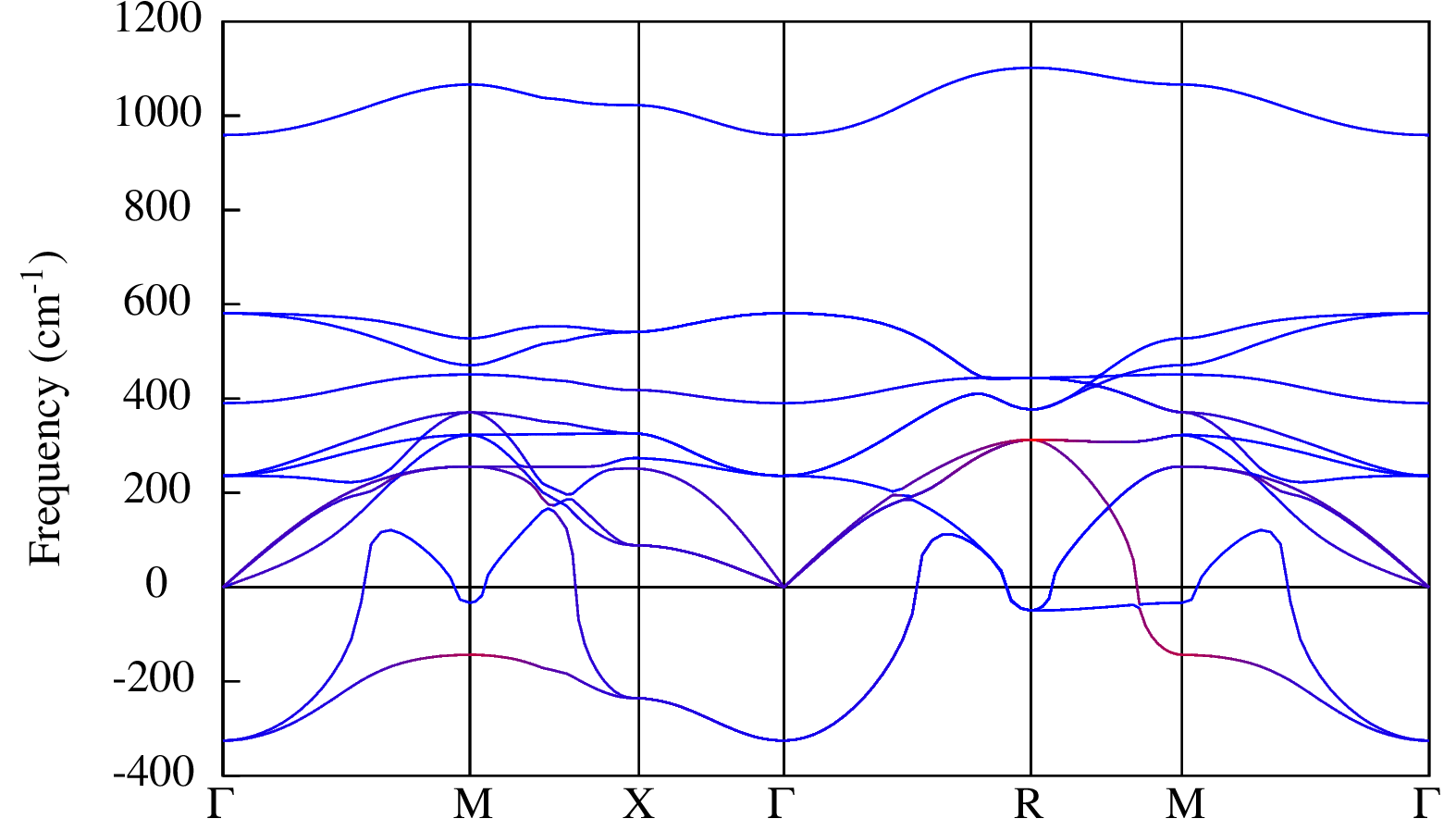} 
\caption{\label{fig:1} The phonon dispersion curve of the cubic phase of WO$_3$, computed with the B1-WC hybrid functional and proper treatment of the dipole-dipole interactions. Lines are colored according to the involvement of W (red) and O (blues) atoms in each mode.}
\end{figure}

We begin our study by considering the high-symmetric parent cubic structure (\textit{$Pm\overline{3}m$}) of WO$_3$. Fig.~\ref{fig:1} shows its phonon dispersion curves calculated using the B1-WC hybrid functional. The high-frequency branches show slightly distinct dispersion than those previously reported in \cite{Hamdi2016}. The previous results were unfortunately affected by a bug in the implementation of the Fourier interpolations technique in the CRYSTAL package while the new curves reported here rely on a proper treatment of the dipole-dipole interactions as implemented in ABINIT (for further information we refer to Fig.~\ref{fig:A1a} in the appendix). 

\begin{table*}
\caption{\label{tab:table1}%
Experimental and calculated lattice parameters (\AA) of different distorted phases of WO$_3$, calculated by LDA, GGA-WC, PBEsol, HSE06 and B1-WC hybrid functionals. Also the PBE results of Yang \textit{et al.}~\cite{Yang2019} are included. Data in parentheses represent relative deviations from the experimental values in percent.
}
\begin{ruledtabular}
\begin{tabular}{lccccccc}
\textrm{Phase}&
\textrm{Exp}&
\multicolumn{1}{l}{\textrm{HSE06     $(\Delta\%)$}}&

\multicolumn{1}{l}{\textrm{B1-WC   $(\Delta\%)$}}&

\multicolumn{1}{l}{\textrm{GGA-WC   $(\Delta\%)$}}&

\multicolumn{1}{l}{\textrm{PBEsol   $(\Delta\%)$}}&

\multicolumn{1}{l}{\textrm{PBE~\cite{Yang2019}} $(\Delta\%)$}&

\multicolumn{1}{l}{\textrm{LDA $(\Delta\%)$}}\\

\colrule
 & 5.29\footnote{Ref~\cite{Howard2002}} &5.29 (+0.0)  & 5.29 (+0.0)   & 5.31 (+0.3)  & 5.30 (+0.1) & 5.37 (+1.5) &5.27 (-0.3)  \\ 
 \textit{P4/nmm}& 3.92 &3.96 (+1.0)  &  3.92 (+0.0)  & 3.92 (+0.0)  & 3.92 (+0.0)& 4.05 (+3.3) &3.87 (-1.1)  \\
\colrule
 & 5.27$^a$  & 5.20 (-1.3) & 5.16 (-2.0)  & 5.26 (-0.1) & 5.24 (-0.5) &- & 5.17 (-1.8) \\
\textit{P4/ncc} & 7.84 & 7.95 (+1.4) &  7.87 (+0.3) & 7.85 (+0.1) & 7.85 (+0.1) &- & 7.74 (-1.2) \\
\colrule
  & 5.27$^a$& 5.28 (+0.1) &5.26  (-0.1) & 5.33  (+1.1) &5.27 (+0.0) &-& 5.23  (-0.7) \\
$P2_1/c$-HT & 5.26 & 5.21 (-0.9) & 5.15 (-2.0)    & 5.31 (+0.9) &5.27 (+0.1) &- &5.16 (-1.9)\\
  & 7.83 &7.70 (-1.6)  &7.61 (-2.8) & 7.72 (-1.4)  & 7.72 (-1.4)  &- & 7.62 (-2.7) 
  \\
  \colrule
 & 7.33\textit{\footnote{Ref~\cite{Vogt1999}}}  &7.46 (+1.7) & 7.28 (-0.6)   & 7.45 (+1.6)  & 7.42 (+1.2)  &7.46 (+1.7) & 7.36 (+0.4)    
\\
\textit{Pbcn} & 7.57& 7.50 (-0.9)  & 7.52 (-0.6) & 7.62 (+0.6)   & 7.59 (+0.2)   &7.61 (+0.5)   & 7.39 (-2.3) 
\\
 & 7.74 &7.69 (-0.6) & 7.68 (-0.7)&  7.76 (+0.2) & 7.76 (+0.2) & 7.98 (+3.1)&7.62 (-1.5) 
\\
 \colrule
 & 7.30$^a$ & 7.43 (+1.7) &  7.37 (+0.9)   & 7.52 (+3.0)  & 7.51 (+2.8)  &   7.49 (+2.6) &7.37 (+0.9) 
\\
\textit{$P2_1/n$} & 7.53 & 7.55 (+0.2) &7.47 (-0.7) &   7.61 (+1.0)  & 7.59 (+0.7) &7.54 (+0.1) &7.39 (-1.8) 
\\
  & 7.69 & 7.62 (-0.9) & 7.54 (-1.9) & 7.67 (-0.2)  &7.66 (-0.3)  &8.02 (+4.2) &7.59 (-1.3) 
\\
\colrule
& 7.30\footnote{Ref~\cite{Diehla15962}} & 7.36 (+0.8) & 7.33 (+0.4)    &7.48 (+2.4)  & 7.44 (+1.9)&7.50 (+2.7) &7.32 (+0.2)\\
$P\overline{1}$ & 7.52 & 7.54 (+0.2) & 7.44 (-1.0)   & 7.62 (+1.3) &7.59 (+0.9) & 7.60 (+1.0)& 7.43 (-1.1) \\
& 7.67 & 7.70 (+0.3)  &  7.61 (-0.7)   & 7.72 (+0.6)  &7.72 (+0.6) &7.98 (+3.8) &7.63 (-0.5)\\

  \colrule
 & 5.27 \footnote{Ref~\cite{Salje1997}}& 5.28 (+0.1) &5.26  (-0.1) & 5.33  (+1.1) &5.27 (+0.0) &-& 5.23  (-0.7) 
\\
$P2_1/c$-LT & 5.15 & 5.21 (+1.1) & 5.15 (+0.0)    & 5.31 (+3.1) &5.27 (+2.3) &- &5.16 (+0.1)
\\
  & 7.66 &7.70 (+0.5)  &7.61 (-0.6) & 7.72 (+0.7)  & 7.72 (+0.7)  &- & 7.62 (-0.5)
\\
\end{tabular}
\end{ruledtabular}
\end{table*}

\begin{table} 
\caption{\label{tab:table2-1}%
Lattice parameters (\AA) of metastable distorted phases of WO$_3$ not observed experimentally, calculated using the LDA, GGA-WC, PBEsol, HSE06 and B1-WC hybrid functionals. Dash symbols
correspond to cases for which the phase is not metastable and the system relaxes back to a higher symmetry.}

\begin{ruledtabular}
\begin{tabular}{lccccc}
\textrm{Phase}&
\multicolumn{1}{l}{\textrm{HSE06}}&

\multicolumn{1}{l}{\textrm{B1-WC}}&

\multicolumn{1}{l}{\textrm{GGA-WC}}&

\multicolumn{1}{l}{\textrm{PBEsol}}&

\multicolumn{1}{l}{\textrm{LDA}}\\

\colrule

\textit{$Pm\overline{3}m$}& 3.78&  3.78 &  3.79&    3.79&3.76  \\
\colrule
 & 5.30 & 5.26 &- & 5.35& 5.25   \\
\textit{P4/mbm} & 3.79 &3.78  &-&3.79& 3.76 \\
\colrule
& 3.88 & 3.85 & 3.85 &3.85& 3.81  \\
$P2_1/m$ &7.43& 7.44 & 7.46& 7.45 & 7.41 \\
  &3.88& 3.85&3.85&3.85&3.81 \\
\colrule
\textit{$R\overline{3}c$}  &5.31& 5.35 & 5.36 & 5.35& 5.29 \\
\colrule
\textit{R3m}&3.82&3.81 &3.81& 3.81&3.78 \\
\colrule
\textit{R3c}& 5.39& 5.35& -& -&5.32  \\
  \colrule
 & 5.32 & 5.32 & -&-& 5.30   \\
\textit{Pnma} &7.44& 7.42 & -&-&7.39 \\
  &5.23&5.22&-&-&5.20\\

\end{tabular}
\end{ruledtabular}
\end{table} 

Imaginary frequencies appear as negative numbers in Fig.~\ref{fig:1} and are associated to phonon structural instabilities of the parent cubic phase.
In fact, the distinct phases of WO$_3$, observed at different temperatures, arise from the individual or combined condensation of the cooperative atomic motions related to these unstable phonon modes. The latter include (i) polar modes (PM) at $\Gamma$ (corresponding to irreducible representation $\Gamma^-_4$), which are related to the motion of W against the O atoms, (ii) anti-polar modes (AP) at the M and X points (M$^-_3$,\ X$^-_5$), which correspond to opposite displacements W against the O atoms from unit cell to unit cell along the [110] and [100] directions, respectively, leading to the displacement of the W atoms from
the center of the octahedral and forming chains with alternating short and long W–O bond lengths, and finally (iii) antiferrodistortive modes (AFD)  at the M and R points (M$^+_3$ and R$^+_4$), which consist in rotation and tilts of the oxygen cages about the central W atom. 

Phonon dispersion curves obtained with different functionals are compared in the appendix. Although they appear very similar, we point out that the  M$^+_3$ AFD mode is slightly unstable using hybrid and LDA functionals while it is slightly stable using GGA functionals. 

In what follows, we will consider all the experimentally observed phases 
as well as six additional metastable phases arising all from distinct combinations of the unstable modes to gain more insight into the energy landscape of WO$_3$. Full structural optimizations of the lattice parameters and atomic positions have then been performed for each phase using different exchange-correlation functionals.  

\subsection{\label{sec:level2}Lattice parameters}

In Table~\ref{tab:table1}, we compare to experiment the calculated lattice parameters obtained with the HSE06 and B1-WC hybrid functionals, the WC and PBEsol GGA functionals, the LDA functional as well as the results relying on the PBE GGA functional as reported recently by Yang \textit{et al.}~\cite{Yang2019}. 
%
In line with what was previously reported by Hamdi \textit{et al.}~\cite{Hamdi2016}, starting from the experimental structure with the $Pc$ space group and performing full structural relaxations, the system relaxes back to a $P2_1/c$ phase of higher symmetry using all considered functionals. Although a $P2_1/c$ phase has only been experimentally observed at about 1000 K~\cite{Howard2002}, its structural parameters are very similar to those of the $Pc$ phase observed at about 200 K~\cite{Salje1997} (the only difference is a small polar distortion as further discussed in the next Section). In Table~\ref{tab:table1}, the lattice parameters of the relaxed $P2_1/c$ phase are therefore compared both to those of the high-temperature $P2_1/c$ phase ($P2_1/c$-HT) and low-temperature $P_c$ phase to which we also refer to as $P2_1/c$-LT hereafter.


For the high-temperature $P4/nmm$ phase, all functionals provide a rather fair agreement with experimental data although PBE results of Yang \textit{et al.} show a significant overestimation (also for other phases). Both GGA-WC and PBEsol provide also a good prediction of the  lattice constants for the $P4/ncc$ phase.  It should be noticed however that the experiments on these high-T phases are performed above 1000 K ~\cite{Howard2002} while the calculations correspond to 0 K. It is therefore possible that this good agreement arises from a cancellation of errors coming, on the one hand, from neglecting the thermal expansion and, on the other hand, from a natural propensity of these GGA functionals  to overestimate the lattice parameters of WO$_3$. In line with that suggestion, the GGA-WC and PBEsol are indeed in better agreement with the $P2_1/c$-HT phase and their overestimate of lattice parameters tends to increase for the low-temperature phases. On the contrary, the LDA tends to systematically underestimate the lattice constants as expected and provides a good agreement for the $P2_1/c$-LT phase. HSE06 provides good estimates although with a tendency to slightly overestimate the volume of the low-temperature phases. B1-WC is confirmed to be the one providing the best overall estimates for all phases, and gives good description of the $P2_1/c$-LT and $P2_1/n$ phases under investigation here. The computed lattice parameters of few additional metastable phases, not observed experimentally, are reported in Table \ref{tab:table2-1}.

\begin{figure*}
\centering
\includegraphics[width=1\textwidth]{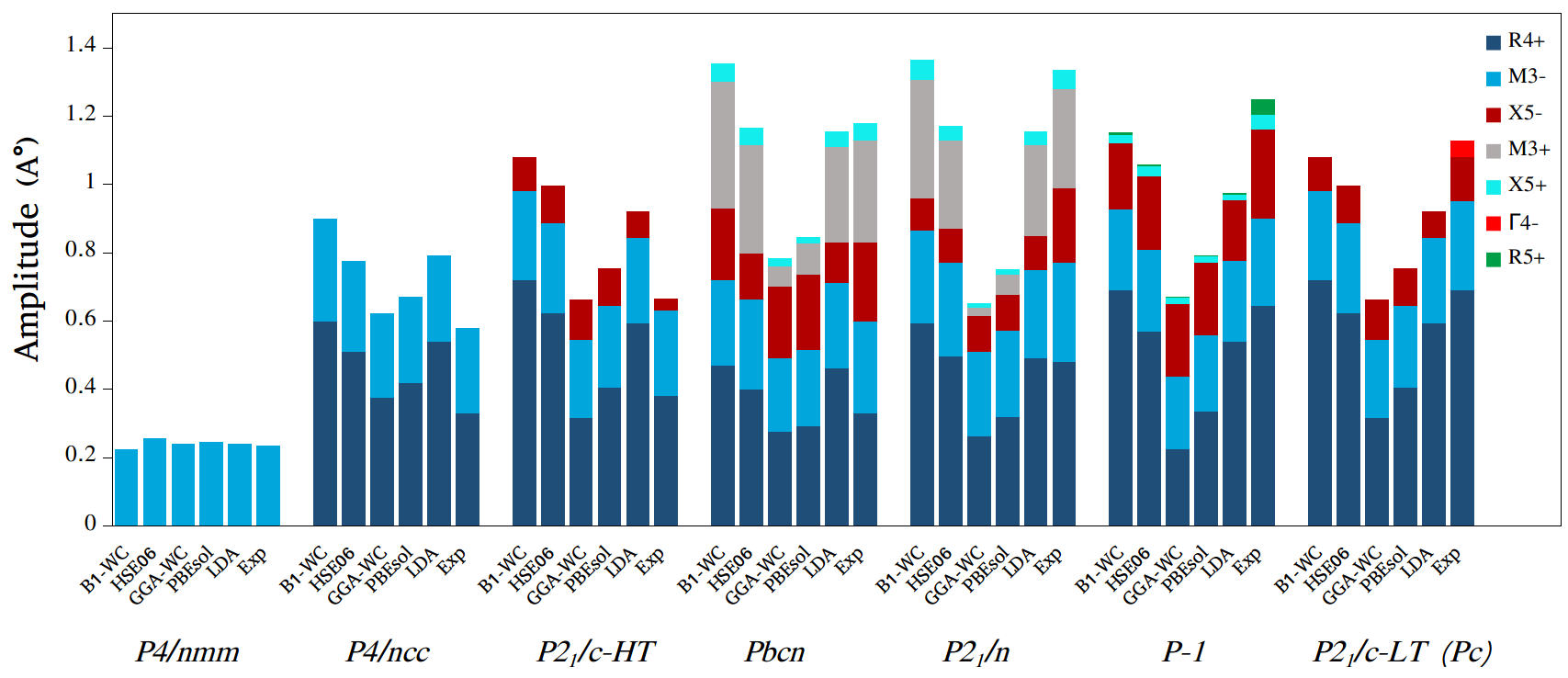}
\caption{\label{fig:2}Symmetry adapted mode decomposition of distorted WO$_3$ phases that are observed experimentally. Comparison between B1-WC, HSE06, GGA-WC, PBEsol, LDA and experimental results~\cite{Howard2002,Vogt1999,Salje1997}. The modes with an amplitude lower than 0.04 Angstrom are not shown.}
\end{figure*}

\begin{figure*}
\includegraphics[width=1\textwidth]{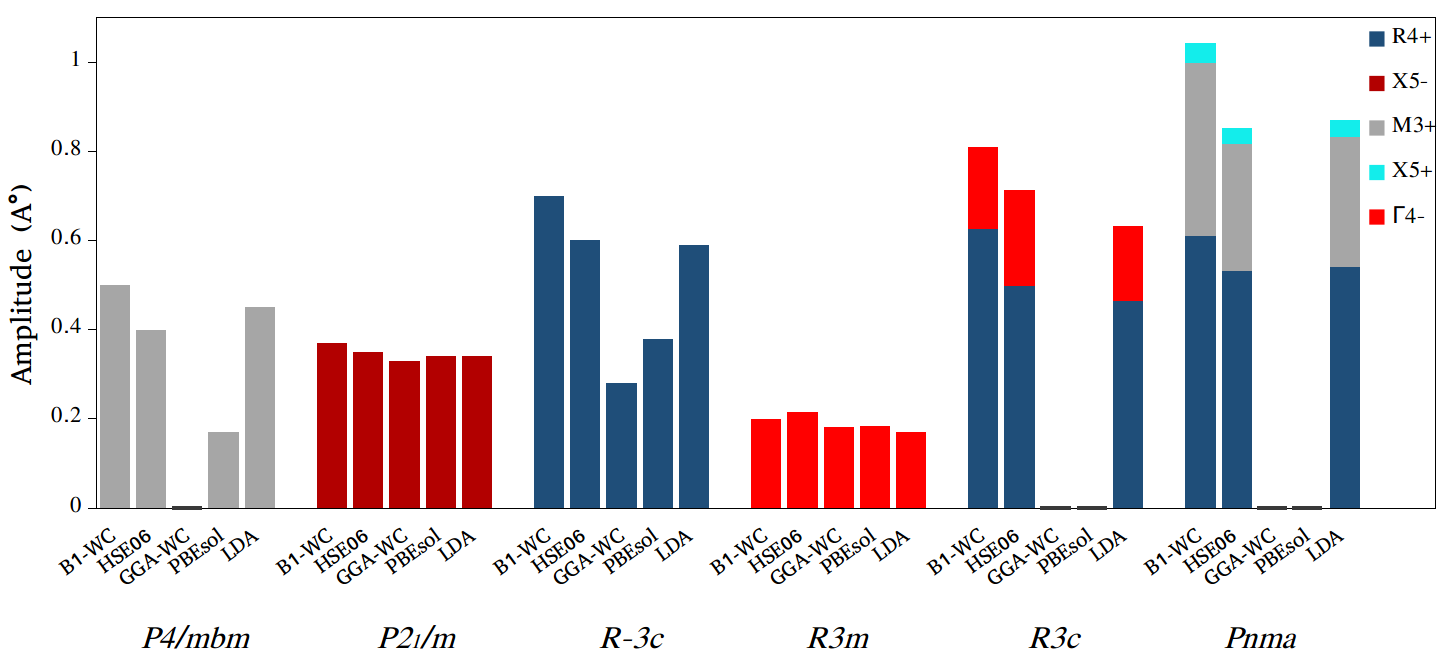}
\caption{\label{fig:3}The symmetry adapted mode decomposition of distorted WO$_3$ phases which are not observed experimentally. Comparison of B1-WC, HSE06, GGA-WC, PBEsol and LDA. The modes with an amplitude lower than 0.04 Angstrom are not shown. The black lines correspond to the cases where the phases are not condensed.}
\end{figure*}

\subsection{\label{sec:level3}Atomic distortion} 

In this section, we analyze further the structural distortions of each phase with respect to the reference cubic phase by projecting  the internal atomic displacements on symmetry adapted modes of the parent cubic phase using the AMPLIMODE software package~\cite{Orobengoa2009}. The results are presented in Fig.~\ref{fig:2} for the  experimentally observed phases and in Fig. \ref{fig:3} for the few other phases. For the latter, missing phases in GGAs  correspond to cases for which the phase is not metastable and the system relaxes back to a higher symmetry. 

As highlighted in Fig.~\ref{fig:2}, the $P4/nmm$ phase arises from the condensation of the only anti-polar $M^-_3$ mode along the $z$ direction. This mode will remain present in all experimentally observed phases with a rather constant amplitude. The $P4/ncc$ phase includes the additional condensation of the AFD R$^+_4$ mode along the same direction yielding a tilt pattern ($a^0a^0c^-$) in Glazer's notations. 
The $P2_1/c$ phase arises from the addition of AFD R$^+_4$ contributions, yielding a tilt pattern ($a^-a^-c^-$). A secondary antipolar $X_5^-$ mode in the $xy$-plane also appears thanks to an anharmonic coupling between the R$^+_4$ and M$^-_3$ modes~\cite{Bousquet2008, Young2015,C5DT00010F}. In the $Pbcn$ phase, a M$^+_3$ mode emerges giving rise to a distinct tilt pattern $a^0b^+c^-$. The secondary antipolar $X_5^-$ mode in this phase is only along the $y$ direction. Also, another secondary antipolar $X_5^+$ mode along $y$ appears through a coupling between the R$^+_4$ and M$^+_3$ modes. Besides, the antipolar $M^-_3$ mode is along the $z$ direction but with a very slight tilt toward the $x$ direction. This small $x$ component of the $M^-_3$ mode appears through a coupling between the secondary $X_5^+$ and $X_5^-$ modes and the primary R$^+_4$ mode. The $P2_1/n$ phase involves the same modes but combined differently within a tilt pattern $a^-b^+c^-$. Again, the antipolar $M^-_3$ mode is along $z$ with a very small $x$ component. Also, the antipolar $X_5^-$ and $X_5^+$ modes are along $y$ direction but with a very small $z$ component. The $P\bar{1}$ phase shows an out-of-phase octahedral tilting around the second axis yielding to tilt pattern $a^-b^-c^-$. This phase has the same antipolar $M_3^-$ and $X_5^+$ modes as in the $P2_1/n$ phase, but the antipolar $X_5^-$ mode has components in all directions. Finally, the experimental $Pc$ phase appears similar to the $P2_1/c$ phase except for the additional appearance of a polar $\Gamma_4^-$ mode. As previously discussed, our calculations in $Pc$ symmetry always relax back to $P2_1/c$ to which the $Pc$ phase is therefore compared. We insist that all the phases except $Pc$ are non-polar. Yang \textit{et al.}~\cite{Yang2019} report a net dipole moment for the $P2_1/n$ phase which is however inconsistent with the symmetry assignment. Fig.~\ref{fig:3} also compares the results for few additional metastable phases not observed experimentally.

 From Fig.~\ref{fig:2} and Fig.~\ref{fig:3}, it appears that, on the one hand, all functionals  predict rather consistently the appearance and amplitudes of polar and antipolar distortions ($M_3^-$, $X_5^-$, $X_5^+$, $\Gamma_4^-$) while, on the other hand, GGA functionals strongly underestimate the amplitudes of rotations and tilts ($R_4^+$, $M_3^+$), as further discussed in the next Section. In line with that, GGAs do not reproduce some metastable phases and also produce the worst description of the $P2_1/n$ and and $P2_1/c$ phases. Hybrid functionals typically provide a very good description of all phases while LDA gives also reasonable results, significantly better than GGAs.

\begin{figure*}
\centering
 \hspace*{-0.2cm}
\includegraphics[width=0.99\textwidth]{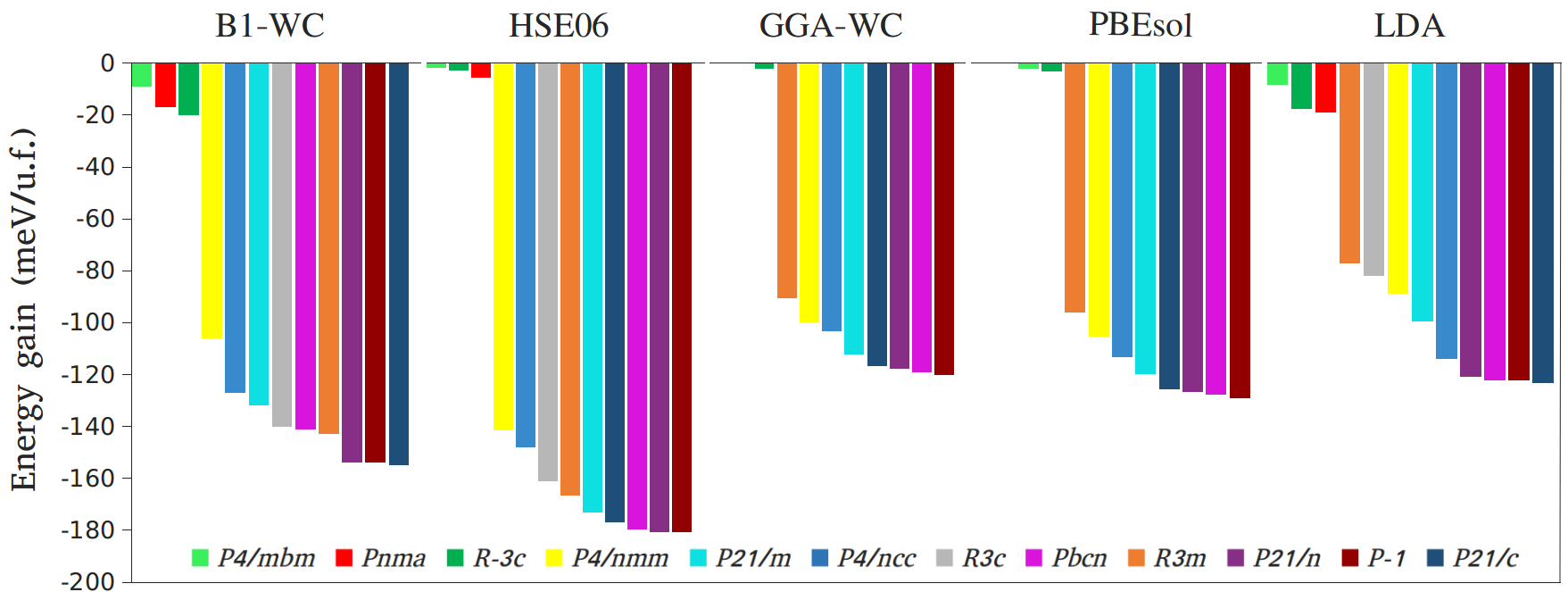}
\caption{\label{fig:4}Calculated energy gain with respect to the cubic phase of different phases of WO$_3$. Comparison between B1-WC, HSE06, GGA-WC, PBEsol and LDA.}
\end{figure*}
\noindent
\noindent
\subsection{\label{sec:level14}Energetics}
Fig.~\ref{fig:4} compares the energies of the various phases under study in comparison to that of the parent cubic phase taken as zero energy reference. The results are shown for  distinct DFT functionals. The energy is reported in meV per four-atoms formula unit (meV/f.u.).

A first observation is that, globally, the gains of energy are slightly larger using the hybrid functionals than the LDA/GGA functionals. This is in line with the slightly larger amplitudes of phonon unstabilities predicted with the hybrid functionals (see the appendix). It is also consistent with the slightly larger amplitudes of distortions obtained with the hybrid functionals, although LDA shows energies more comparable to GGA but amplitudes of distortion closer to hybrid functionals.

Comparing the lowerings of energy produced by the individual condensation of distinct kinds of instabilities (AFD, PM, AP), we observe that the smallest ones are related to the AFD motions ($P4/mbm$, $Pnma$, $R\bar{3}c$ phases). This is compatible with the fact that AFD modes are less unstable than FE or AP modes. The energy lowerings produced by AFD modes are also significantly smaller with GGA functionals that also show slightly smaller AFD instabilities (the $M_3^+$ mode is even stable in GGA-WC). Consequently, some AFD phases are not even appearing as metastable in GGA.  We further notice that the energy lowerings produced by AFD modes are also smaller with HSE06 than B1-WC although in this case the phonon dispersion curves and the amplitudes of distortion are relatively close. Only LDA shows an energetics of AFD comparable to B1-WC.

Regarding the individual condensation of PM and AP modes, although the strongest instability is observed in the dispersion curves for the $\Gamma^-_4$ mode with all functionals, the largest gain of energy arises from the condensation of the AP X$^-_5$  ($P2_1/m$ phase) except with the B1-WC for which the $R3m$ phase resulting from the condensation of $\Gamma^-_4$ is lower in energy. We notice also that while condensation of the AP M$_3^-$ mode ($P4/nmm$ phase) produces a lowering of energy smaller than the polar $\Gamma^-_4$ ($R3m$ phase) with hybrid functionals, the opposite is true for LDA and GGA functionals.

As previously discussed by Hamdi {\it et al.} \cite{Hamdi2016}, the lowest energy phases of WO$_3$ result from the combination of distinct instabilities and anharmonic couplings play an important role in the formation of $Pbcn$, $P2_1/n$, $P\overline{1}$ and $P2_1/c$ phases. For this reason it is not straightforward to clarify the origin of differences in the energy landscape obtained with different functionals. As illustrated in Fig.~\ref{fig:4}, all functionals locate consistently these four phases very close in energy (except for B1-WC that locates the $Pbcn$ phase at slightly higher energy) but nevertheless in a different order for each of them. We notice that only B1-WC and LDA predict the $P2_1/c$ phase as the ground state. Moreover, although not a requirement for internal energies, the energetics of the different phases in B1-WC is in line with the sequence of phase transitions observed experimentally. 

All this highlights that WO$_3$ exhibits a very delicate energy landscape. Choosing the most appropriate functional remains a delicate issue that goes beyond trying to best reproduce cell parameters and atomic distortions. Our study confirms the B1-WC hybrid functional as the most appropriate choice and points the LDA as a reasonable alternative.

\begin{figure}

\includegraphics[width=0.46\textwidth]{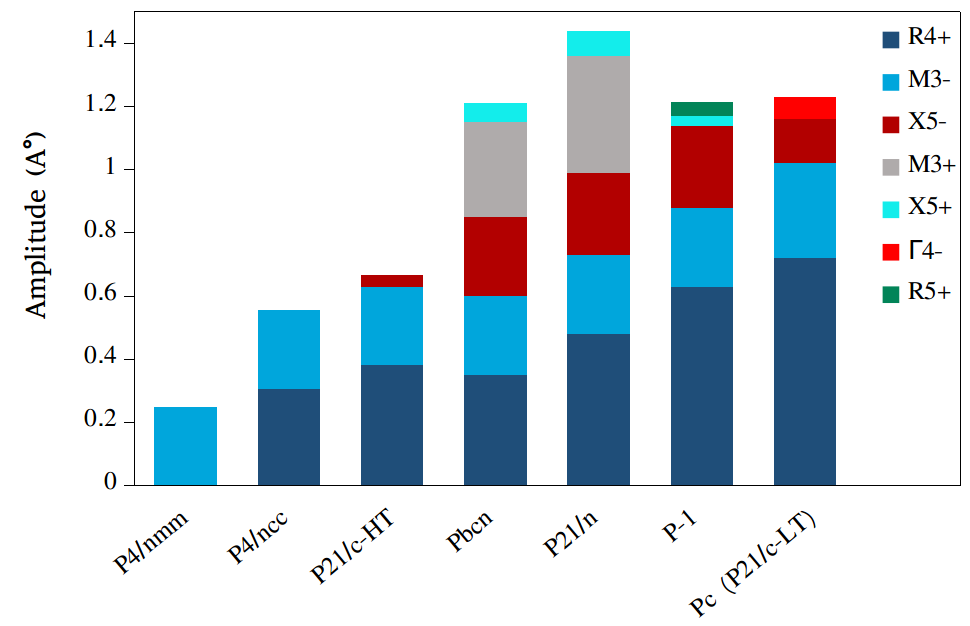}
\caption{\label{fig:fig5}The experimental symmetry adapted mode decomposition of observed phases presented in the order they appear with decreasing temperatures.}
\end{figure}

\subsection{\label{sec:level44}Phase diagram}

In line with its complex energy landscape, WO$_3$ presents a complicated phase diagram which, as mentioned before, is still partly debated. One question is the potential polar nature of its ground state phase: as previously discussed and confirmed by the present study, the $P2_1/c$ phase of stoichiometric WO$_3$ is predicted as the theoretical ground state and does not show any tendency to polar instability so that the experimental $P_c$ assignment, deviating only marginally from the $P2_1/c$ prediction, might be related to extrinsic effects \cite{Bousquet2020}. Then, particularly puzzling is the intermediate $P2_1/c$ phase appearing in a narrow temperature range at high-temperature and re-emerging as the ground state at low temperature.

In Fig. \ref{fig:fig5}, we compare the symmetry-mode analysis of all experimental phases from high to low temperatures. The AP M$_3^-$ is common to all phases and remains roughly constant in amplitude. Then, on cooling, the AFD R$^+_4$ mode appears from the $P4/ncc$ phase and is preserved down to the ground state.  We notice that the Glazer tilt pattern associated to the different phases (i.e. number and orientation of R$^+_4$ modes) is evolving from one phase to another. In the $Pbcn$ and $P2_1/n$ phases there is the emergence of an additional AFD M$_3^+$ distortion which however compete with the R$^+_4$ one \cite{Hamdi2016} and disappears in the $P\bar{1}$ and $P2_1/c$-LT phases. 

The main difference between the high- and low-temperature $P2_1/c$ phases is the amplitude AFD R$^+_4$ mode. It might happen that at high-temperature at which the R$^+_4$ distortion are reduced, the M$_3^+$ distortion can appear to produce intermediate phases while, due to mode competition, the latter disappear when the R$^+_4$ distortion becomes larger at low temperatures.


Our calculations show that the  $Pbcn$, $P2_1/n$, $P\overline{1}$ and $P2_1/c$-LT phases are very close in energy. This might contribute to explain the origin of the very complex and unusual phase diagram of WO$_3$. It suggests also that distinct phases could coexist and/or could be stabilized under slight variation of the experimental conditions.

\noindent
\section{Born effective charges and dielectric constant}

We consider now the Born effective charges ($Z^*$) and dielectric tensor ($\epsilon^{\infty}$). All results in the main text are obtained using the LDA functional. Since the optimized LDA lattice parameters slightly underestimate  the experimental values, for better comparison with experiment, we decided to work with phases for which (i) lattice parameters are fixed at the experimental values for $P2_1/n$ phase and at the optimized B1-WC values for $P2_1/c$ phase and (ii) atomic positions are fully relaxed under that constraint. In practice, the deviations of atomic positions remain negligible compared to the structures fully relaxed with B1-WC (see Tab.~\ref{tab:tableS44}). Further comparison of the $Z^*$ and $\epsilon^{\infty}$ computed in LDA and B1-WC functionals are also provided in the appendix (Tables A2 and A3), showing close agreement.

In Tab.~\ref{tab:table2} the Born effective charge tensors ($Z^*$) of WO$_3$ are presented for the eight nonequivalent  atoms in the $P2_1/n$ phase, and the four nonequivalent  atoms in the $P2_1/c$ phase. Some $Z^*$ are anomalously large in comparison to the nominal atomic charges ($+6 e$, $-2 e$ for W and O atoms, respectively) and these values can be linked to the dynamical changes in the hybridization between the $2p$ orbitals of the oxygen and the $5d$ orbitals of the W atoms. The {\it main} values of the symmetric part of these tensors have also been reported for easier comparison of the different phases and atoms. 

 In the cubic structure of WO$_3$, the WO$_6$-octahedra are undistorted and no symmetry breaking is observed. This implies the existence of diagonal, isotropic Born effective charge tensors: $+13.47 e$ for W, $-9.79 e$ for O$_{||}$, and $-1.8 e$ for O$_{\perp}$ consistent with those previously reported in Ref. \cite{PhysRevB.56.983}. By distorting the cubic environment, the off-diagonal values of $Z^*$ increase and the main values of $Z^*$ decrease with respect to the cubic phase, as previously observed in perovskites \cite{PhysRevB.51.6765, PhysRevB.58.6224}. In both the $P2_1/c$ and $P2_1/n$ phases the off-diagonal terms of $Z^*$ are small, with the only exceptions the O$_1$ and O$_2$ atoms in the basal plane of the $P2_1/c$ crystal. This is due to the fact that $Z^*$ is presented in the cartesian basis while the unit cell basis of $P2_1/c$ is rotated by 45 degrees around the $z$ axis. 
 
We believe that the significant differences between the diagonal components of $Z^*$($W_1$) (and $Z^*$($W_2$)) for the $P2_1/n$ phase can be associated with the distance between two consecutive tungsten atoms in W-O-W chains, which are 3.70, 3.80 and 3.84 \r{A} along the $x$, $y$ and $z$ directions, respectively. Therefore, by decreasing the interatomic distance the diagonal components of $Z^*$($W_1$) and $Z^*$($W_2$) will increase. Furthermore, the same trend exists when the long-short bond alternation of W-O is decreased. Although the same pattern is observed for the $P2_1/c$ phase, the differences between the diagonal components of $Z^*$($W_1$) are smaller than for the $P2_1/n$ phase.

 \begin{table*}
\caption{\label{tab:table2}
Calculated (LDA) Born effective charge tensors ($Z^*$) and their main values (between parentheses) of the atoms of the asymmetric units (see notations in Tab.~\ref{tab:tableS44} in the appendix) of the $P2_1/n$ and $P2_1/c$ phases of WO$_3$. 
}
\begin{ruledtabular}
\begin{tabular}{cccc}
\textbf{$P2_1/n$}&&
&\\
\colrule
$Z^*$($W_1$)   & $Z^*$($O_1$) & $Z^*$($O_3$) & $Z^*$($O_5$)  \\
   $\begin{pmatrix}
  10.79 & -1.15 & 0.64\\ 
  0.75 & 8.55 & -0.37\\ 
  -0.49 & -0.30 & 7.79
\end{pmatrix}$                      &    $\begin{pmatrix}
  -8.00 & 0.93 & -0.94\\ 
  0.34 & -1.25 & 0.08\\ 
  -0.44 & 0.10 & -1.12
\end{pmatrix}$   &    
   $\begin{pmatrix}
  -1.43 & 0.09 & -0.10\\ 
  0.16 & -6.22 & -0.69\\ 
  0.05 & -0.28 & -1.02
\end{pmatrix}$    &  $\begin{pmatrix}
  -1.35 & 0.03& -0.05\\ 
  -0.01 & -1.05 & -0.45\\ 
  -0.05 & -0.14 & -5.65
\end{pmatrix}$     \\
& & & \\

 $\begin{pmatrix}
  10.81 & 8.65 & 7.66\\    
\end{pmatrix}$& $\begin{pmatrix}
  -8.12 &-1.19 & -1.04\\    
\end{pmatrix}$&  $\begin{pmatrix}
  -6.26 &-1.42 & -0.97\\    
\end{pmatrix}$& $\begin{pmatrix}
  -5.66 &-1.34 & -1.03\\    
\end{pmatrix}$\\
& & & \\
$Z^*$($W_2$)   & $Z^*$($O_2$) & $Z^*$($O_4$) & $Z^*$($O_6$)  \\
 $\begin{pmatrix}
  10.80 & 1.17 & -0.78\\ 
  -0.78 & 8.62 & -0.66\\ 
  0.36 & -0.04 & 7.82
\end{pmatrix}$                  &   $\begin{pmatrix}
   -8.01 & 0.93 & 1.00\\ 
  0.35 & -1.25 & -0.08\\ 
  0.39 & -0.10 & -1.14
\end{pmatrix}$    &    
 $\begin{pmatrix}
  -1.43 &-0.10 & 0.08\\ 
  -0.16 & -6.28 & -0.75\\ 
  -0.05 & -0.27 & -1.03
\end{pmatrix}$   &  $\begin{pmatrix}
  -1.35 & 0.01 & 0.13\\ 
  -0.03 & -1.07 & 0.47\\ 
  0.23 & 0.19 & -5.63
\end{pmatrix}$    \\ 
& & & \\

 $\begin{pmatrix}
  10.83 & 8.72 & 7.68\\    
\end{pmatrix}$&  $\begin{pmatrix}
  -8.14 &-1.19 & -1.06\\    
\end{pmatrix}$& $\begin{pmatrix}
  -6.33&-1.39 & -1.01\\    
\end{pmatrix}$ &$\begin{pmatrix}
  -5.59 &-1.22 & -1.18\\    
\end{pmatrix}$   \\
& & & \\

\colrule
\textbf{$P2_1/c$}&
& &\\
\colrule
$Z^*$($W_1$)   & $Z^*$($O_1$) & $Z^*$($O_2$) & $Z^*$($O_3$)  \\
$\begin{pmatrix}
  10.38 & -1.07 & -0.15\\ 
  0.99 & 9.40 & 0.95\\ 
  -0.17 & -0.89 & 7.74
\end{pmatrix}$              &     $\begin{pmatrix}
  -4.57 & 3.05 & -0.66\\ 
  2.92 & -4.13 & 0.54\\ 
  -0.24 & 0.23 & -1.06
\end{pmatrix}$                    &    
 $\begin{pmatrix}
  -4.62 & 2.90 & 0.65\\ 
  2.89 & -4.01 & -0.57\\ 
  0.23 & -0.23 & -1.11
\end{pmatrix}$                &$\begin{pmatrix}
  -1.19 & 0.00 & 0.16\\ 
  -0.03 & -1.25 & 0.21\\ 
  0.18 & 0.45 & -5.56
\end{pmatrix}$          \\
& & & \\

 $\begin{pmatrix}
  10.39 & 9.39 & 7.72\\    
\end{pmatrix}$  & $\begin{pmatrix}
  -7.46 &-1.36 & -1.01\\    
\end{pmatrix}$   & $\begin{pmatrix}
  -7.17 &-1.40 & -1.16\\    
\end{pmatrix}$ & $\begin{pmatrix}
  -5.66 &-1.34 & -1.04\\
  \end{pmatrix}$     

\end{tabular}
\end{ruledtabular}
\end{table*}
\begin{table}
\caption{\label{tab:table3}%
Optical dielectric tensor (${\varepsilon }^{\infty }_{ij}$) of the $P2_1/n$ and $P2_1/c$ phases of WO$_3$ within LDA.
}
\begin{ruledtabular}
\begin{tabular}{cc}
\textrm{$P2_1/n$}&
\textrm{$P2_1/c$} \\

   $\begin{pmatrix}
  7.05 & 0 & -0.02\\ 
  0 & 5.83 & 0\\ 
  -0.02 & 0 & 5.35
\end{pmatrix}$ 
&  $\begin{pmatrix}
  6.95 & 0 & -0.08\\ 
  0 & 6.50 & 0\\ 
  -0.08 & 0 & 5.44
\end{pmatrix}$                             
\end{tabular}
\end{ruledtabular}
\end{table}

\begin{table}\caption{\label{tab:table4}%
Calculated (LDA) and experimental refractive indices ($n_i$) of the $P2_1/n$ and $P2_1/c$ phases of WO$_3$
}
\begin{ruledtabular}
\begin{tabular}{llll}
\textit{Phase}&
\textrm{$n_1$}&
\textrm{$n_2$}&
\textrm{$n_3$} \\
\colrule
$P2_1/n$ (EXP)~\cite{Sawada1959} & 2.70$\pm$0.035 & 2.37$\pm$0.035 & 2.28$\pm$0.035 \\
$P2_1/n$ (LDA) & 2.65 & 2.41 & 2.31 \\

$P2_1/c$ (LDA) & 2.63 & 2.54 & 2.33 \\

\end{tabular}
\end{ruledtabular}
\end{table}
The optical dielectric tensor (${\varepsilon }^{\infty }_{ij}$), and related refractive indices ($n_i$) are reported in Tab.~\ref{tab:table3} and Tab.~\ref{tab:table4}. The experimental values~\cite{Sawada1959} of the refractive indices measured at room temperature are also reported for comparison. The results for the room temperature $P2_1/n$ phase show a good agreement with the experimental data. Our calculations indicate a high index of refraction for both the $P2_1/n$ and $P2_1/c$ phases. This is due to the large local field in these two phases of WO$_3$. Also, the birefringence ($\Delta n$) is larger in the $P2_1/n$ phase (0.23, 0.34 and 0.1 for $n_1$-$n_2$, $n_1$-$n_3$ and $n_2$-$n_3$, respectively) than in the $P2_1/c$ phase \cite{Santato2001}. 

For comparison, we also computed the optical dielectric tensor ${\varepsilon }^{\infty }_{ij}$, and Born effective charges $Z^*$ of the ground state $P2_1/c$ phase using the B1-WC and HSE06 hybrid functional. The results are presented in the appendix Tab.~\ref{tab:s1},~\ref{tab:s2}. We can conclude that LDA results are in reasonable agreement with the B1-WC results. However, LDA usually overestimates the absolute value of ${\varepsilon }^{\infty }$, which is due to the underestimation of the electronic bandgap in line with the lack of polarization dependence of this local exchange-correlation functional~\cite{PhysRevLett.74.4035}.

\section{Dynamical properties}

We focus now on the dynamical properties of the room-temperature $P2_1/n$ phase and ground-state $P2_1/c$ phase of WO$_3$. In each case, we first discuss the phonons at the zone center ($\Gamma$ point), reporting both Raman and infra-red spectra, and also the full phonon dispersion curves. As in the previous Section, all results in the main text have been obtained in LDA. Further comparison of LDA and B1-WC frequencies are reported in the appendix (Table A4), showing very close agreement regarding both phonon frequencies and eigenvectors. 

 \subsection{\label{andlo}The room temperature $P2_1/n$ phase} 
 
\subsubsection{Irreducible representations at $\Gamma$}
For the non-polar $P2_1/n$ structure, which belongs to the $C^5_{2h}$ point group, there are 32 atoms in the primitive cell and so 96 phonon modes at each $k$-point. The zone-center phonons can be classified according to the irreducible representations of this point group as $$\Gamma_{phonons} = 24 A_g \oplus 24 A_u \oplus 24 B_g \oplus 24 B_u .$$ These include: (i)  three acoustic modes ($\Gamma_{acoustic}= A_{u}+ 2B_{u}$), (ii) 48 Raman active modes ($A_g$ and $B_g$) and (iii) 45 infrared (IR) active modes ($A_u$ and $B_u$). 

\subsubsection{Infrared active modes}

\begin{table}
\caption{\label{tab:table5}
The TO phonon frequencies of the infrared active modes of the $P2_1/n$ room temperature phase within LDA in cm$^{-1}$. The values in brackets correspond to the experimental measurements reported in \cite{Salje1975} }
\begin{ruledtabular}
\begin{tabular}{llll}
\textrm{Irrep}&
\textrm{LDA [Exp]}&
\textrm{Irrep}& \textrm{LDA [Exp]} \\
\colrule
$A_u$(1) & 46& $B_u$(1) & 126
\\ 
$A_u$(2) & 57&$B_u$(2) &  194
\\ 
$A_u$(3) & 74& $B_u$(3) & 203
 \\ 
$A_u$(4) &  108&$B_u$(4) & 221
\\ 
$A_u$(5) & 134&$B_u$(5) &235
\\ 
$A_u$(6) & 206&$B_u$(6) &  265
 \\ 
$A_u$(7) & 226 [230]& $B_u$(7) & 280
 \\ 
$A_u$(8) & 257& $B_u$(8) & 283 [285]
\\ 
$A_u$(9) & 266&$B_u$(9) & 303
\\ 
$A_u$(10) & 272& $B_u$(10) & 310 [310]
 \\ 
$A_u$(11) & 308& $B_u$(11) &315
 \\ 
$A_u$(12) & 326& $B_u$(12) &332 [335]
 \\ 
$A_u$(13) & 333& $B_u$(13) & 344
\\ 
$A_u$(14) & 346& $B_u$(14) & 362
\\ 
$A_u$(15) & 362 [370]& $B_u$(15) &379
\\ 
$A_u$(16) & 425& $B_u$(16) &  407
\\ 
$A_u$(17) &  446& $B_u$(17) & 625 [665]
\\ 
$A_u$(18) &609& $B_u$(18) & 756
\\ 
$A_u$(19) & 695& $B_u$(19) & 776
 \\ 
$A_u$(20) &  699& $B_u$(20) & 847 [825]
 \\ 
$A_u$(21) & 766& $B_u$(21) &1042 
\\ 
$A_u$(22) & 1010 [920]& $B_u$(22) & 1048 
\\ 
$A_u$(23) & 1015&-&-
\\
\end{tabular}
\end{ruledtabular}
\end{table}

{\it TO modes --} The calculated TO frequencies of the IR active phonons are listed in Tab.~\ref{tab:table5}, besides the measured experimental data (in brackets) from Ref. \cite{Salje1975} and their possible assignments. In fact, the assignment of the IR active modes remains still experimentally unexplored. Thus, we assigned each experimental frequency to the mode with largest mode effective charge in the same frequency range (see Tab.~\ref{tab:s5} in the appendix). It can be observed that the experimental values and our calculated frequencies are overall in good agreement.

{\it LO modes --}The IR active LO modes take different frequencies when approaching $\Gamma$ from different directions due to the low symmetry of the  $P2_1/n$ structure. In Tab.~\ref{tab:s5} in the appendix, the LO frequencies of the IR active modes along different directions are reported together with related TO frequencies and mode effective charges. Since there is no one-to-one correspondence, the mapping between LO and TO modes is done by comparing the eigenvectors and looking for the maximum overlap.

The polarities of the $A_u$ phonon modes are along the $y$ direction, thus, they do not show any LO-TO splitting along the $\Gamma-Y$ and $\Gamma-B$ directions. On the other hand, some $A_u$ modes experience large shifts of their LO frequencies along the $[0 1 0]$ direction (with a width of 45 cm$^{-1}$ for $A_u$(8), 48 cm$^{-1}$ for $A_u$(15) and 262 cm$^{-1}$ for $A_u$(19)). Along the $[0 1 1]$ direction, although there is a component parallel to the polarity, no significant shift of LO frequencies is found for $A_u$ modes. The $A_u$(15) and $A_u$(19) modes experience a strong LO-TO shifting along $[1 1 0]$ (44 cm$^{-1}$ and 263 cm$^{-1}$, respectively). Along the $[1 1 1]$ direction the $A_u$(7) mode and again the $A_u$(19) mode show giant shifts of 171 cm$^{-1}$ and 268 cm$^{-1}$, respectively. In fact, all these giant shifts correspond to  modes with very large mode effective charges that are associated with the opposite motion of W and O atoms along the $y$ direction.
\begin{table}
\caption{\label{tab:table7}
Phonon frequencies of Raman active modes of the $P2_1/n$ room temperature phase within LDA in cm$^{-1}$.}
\begin{ruledtabular}
\begin{tabular}{cccc}
\textrm{Irrep}&
\multicolumn{1}{c}{\textrm{LDA}}&
\multicolumn{1}{c}{\textrm{Irrep}}& \textrm{LDA} \\
\colrule
$A_g$(1) &48 & $B_g$(1) &46 \\
$A_g$(2) & 53 & $B_g$(2) &75\\
$A_g$(3) & 59 & $B_g$(3) &81 \\
$A_g$(4) & 72 & $B_g$(4) &178 \\
$A_g$(5) &79  & $B_g$(5) & 186\\
$A_g$(6) & 117 & $B_g$(6) &202 \\
$A_g$(7) & 159 & $B_g$(7) & 218\\
$A_g$(8) &206  & $B_g$(8) & 249\\
$A_g$(9) & 214 & $B_g$(9) & 255\\
$A_g$(10) & 269 & $B_g$(10) & 291\\
$A_g$(11) & 271 & $B_g$(11) &312 \\
$A_g$(12) & 304 & $B_g$(12) & 336\\
$A_g$(13) & 321 & $B_g$(13) &354\\
$A_g$(14) & 329 & $B_g$(14) &359 \\
$A_g$(15) & 342 & $B_g$(15) & 363\\
$A_g$(16) &419  & $B_g$(16) &393 \\
$A_g$(17) & 434 & $B_g$(17) &430\\
$A_g$(18) &443  & $B_g$(18) &433 \\
$A_g$(19) & 589 & $B_g$(19) &608 \\
$A_g$(20) & 689 & $B_g$(20) & 738\\
$A_g$(21) & 704 & $B_g$(21) &756 \\
$A_g$(22) & 766 & $B_g$(22) & 842\\
$A_g$(23) & 809 & $B_g$(23) &1003\\
$A_g$(24) & 1051 & $B_g$(24) & 1079\\
\end{tabular}
\end{ruledtabular}
\end{table}
As illustrated in Table~\ref{tab:s5}, the $B_u$ modes have their polarity in the $xz$-plane. The two largest LO-TO splittings of 334 and 275 cm$^{-1}$ are observed for the $B_u$(17) and $B_u$(1) modes respectively and are related to the opposite motion of W and O atoms along $x$. The mode effective charges of the $B_u$(17) and $B_u$(1) modes are 18.1$e$ and 19.5$e$, respectively, along the $x$ direction, and almost zero along $z$, confirming that these modes are mostly polarized along the $x$ direction. Note that this point is in line with the considerably large (xx) components of the $Z^*$($W_1$), $Z^*$($W_2$), $Z^*$($O_1$) and $Z^*$($O_2$) tensors of the $P2_1/n$ phase. 
\begin{figure} 
\hspace*{-0.35cm}
\includegraphics[width=0.51\textwidth]{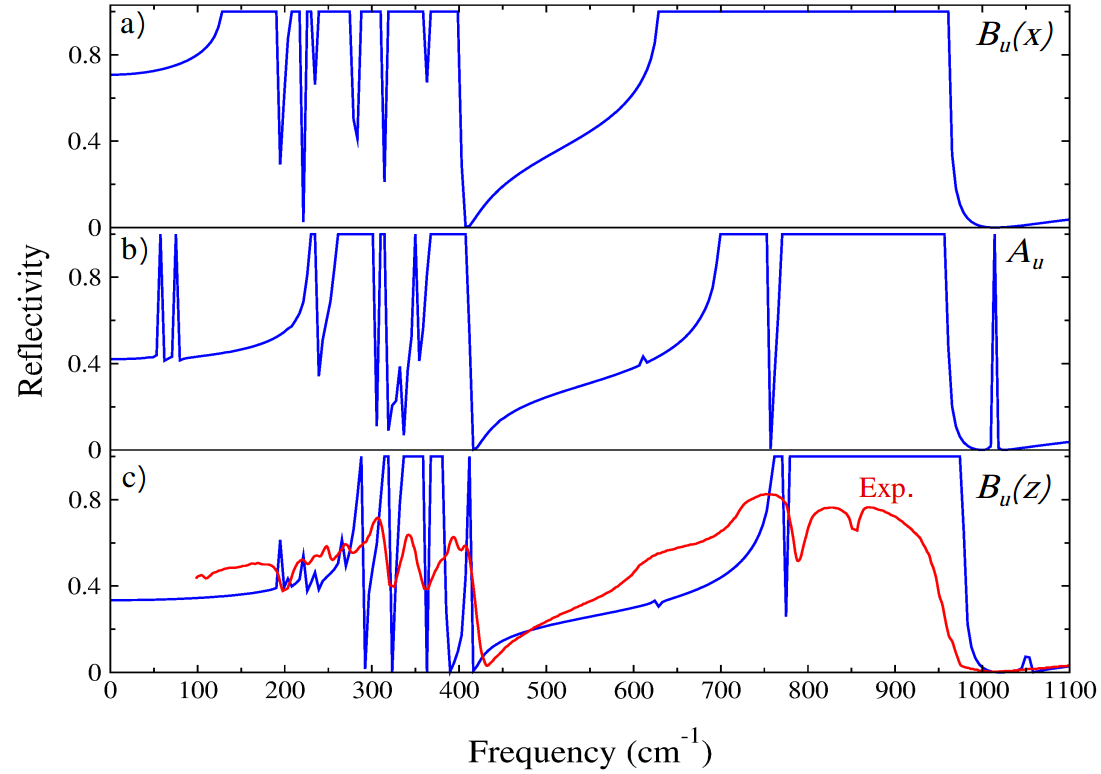}

\caption{\label{fig:fig6}The calculated infrared reflectivity of the $P2_1/n$ phase: (a) $B_u$ with the polarity along $x$, (b) $A_u$, (c) $B_u$ with the polarity along $z$, also the experimental reflectivity spectrum of the $[001]$ crystal surface of WO$_3$ measured at room temperature in the spectral range from 50 to 1200 cm$^{-1}$, taken from Ref.~\cite{Gabrusenoks2001}.}
\end{figure}
\begin{table*}
\caption{\label{tab:table8}
The comparison between calculated Raman modes of WO$_3$ $P2_1/n$ phase within the LDA and available experimental data. The bold fonts represent the frequencies of Raman lines in the calculated Raman spectrum of Fig.~\ref{fig:fig8}.}
\begin{ruledtabular}
\begin{tabular}{ccccccc}
\textbf{Irrep} & \textbf{LDA} &  \textbf{Expt.\cite{Daniel1987}} & \textbf{Expt.\cite{Thummavichai2018}}& \textbf{Expt.\cite{Raul2013}}& \textbf{Expt.\cite{Salje1975}} & \textbf{Expt.\cite{Santato2001}} \\ 
\colrule
$A_g$(1) & 48 &  34& -& - & 33-$A_g$ & - \\
$A_g$(3) & 59& 61& -& 63 & 60 & - \\
$A_g$(4)& 72& 71& -& 71 &73 & - \\
$A_g$(5)& 79 &  93& -& - &93 & - \\
$A_g$(6)& \textbf{117} &134& 137& 136 & 133-$A_g$ & - \\
$A_g$(7) & 159 &187& 187& 187 &- & - \\
$A_g$(9)& 214 & 218& -& 221 & - & - \\
$A_g$(11) & \textbf{271} &  273& 275& 273 & 275 & 272 \\
$A_g$(13) & \textbf{321} &  327& 329& 328 & 330 & 322 \\
$B_g$(16) &\textbf{393}& -& -& 376 & - & - \\
$A_g$(17) & \textbf{434}&   434& -& 436 & - & 430 \\
$A_g$(18) &\textbf{443 }& -& -& - & - & 447 \\
$A_g$(20) & \textbf{689} &  715& 718& 717 & 719-$A_g$ & 709 \\
$A_g$(22) & \textbf{766} &   807& 808& 808 & 808-$A_g$ & 807 \\
\end{tabular}
\end{ruledtabular}
\end{table*}
\begin{figure}
\hspace*{0 cm}
\includegraphics[width=0.48\textwidth]{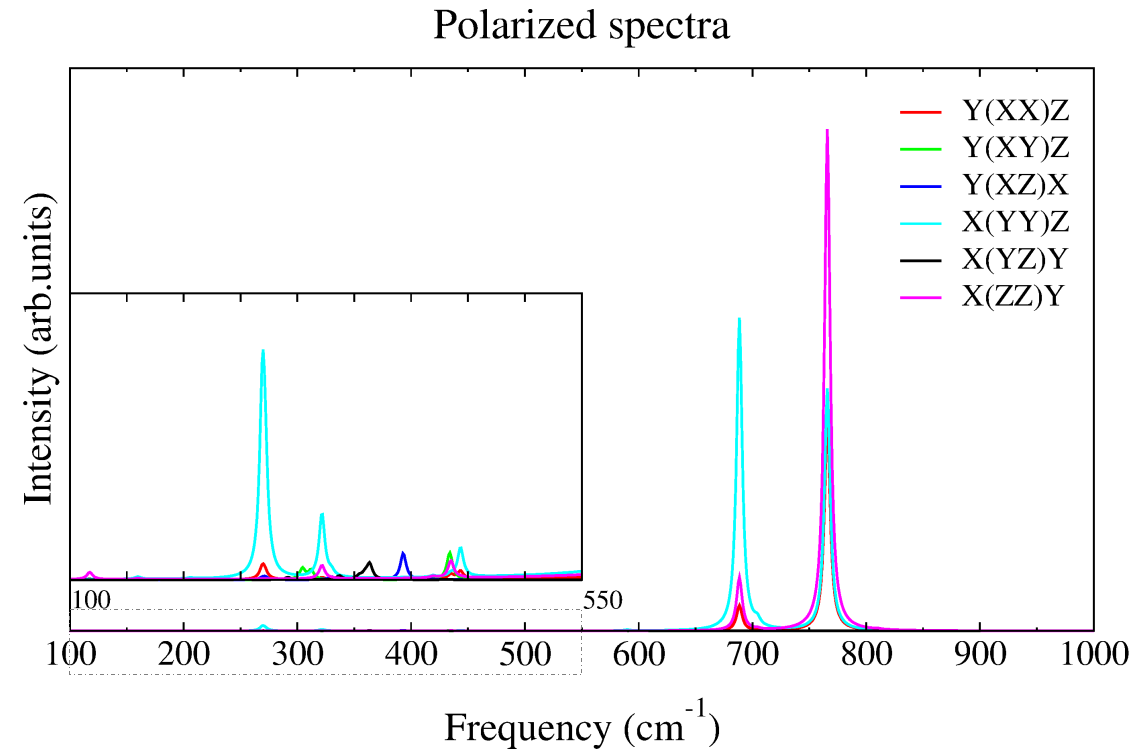} 

\caption{\label{fig:fig7}The polarized Raman spectra of the $P2_1/n$ phase, calculated within the LDA (for visualization purposes the low frequency range between 100-550 cm$^{-1}$ is magnified).}
\end{figure}

{\it IR reflectivity spectra --} The infrared reflectivity spectra are also calculated at normal incidence for all three Cartesian directions, showing contributions of $B_u$ (along $x$), $A_u$ (along $y$) and of $B_u$ (along $z$) modes respectively. The experimental reflectivity spectrum of the $[001]$ crystal surface of WO$_3$ measured by Gabrusenoksis \textit{et al.}~\cite{Gabrusenoks2001} is also presented in panel $(c)$ of Fig.~\ref{fig:fig6}. In our calculations, the reflectivity spectra saturate to unity because our formalism neglects damping. In spite of that, a reasonably good agreement between the theoretical and experimental reflectivity spectrum is observed. However it seems that the small feature at 850 cm$^{-1}$ corresponding to the $B_u(20)$ mode is missed in the theoretical spectrum. This mode is correctly predicted by the calculation but is predicted with a negligible mode charge.

\subsubsection{Raman active modes}

 {\it Raman modes --} Tab.~\ref{tab:table7} shows the calculated frequencies of Raman active phonons classified according to their symmetry. To the best of our knowledge, the only theoretical study of the Raman spectrum of the $P2_1/n$ phase of WO$_3$ so far is done by Yang \textit{et al.}~\cite{Yang2019}. In contrast to both experimental data and our calculations, they report three dominant Raman peaks in the high-frequency range, located around 610, 650, 850 cm$^{-1}$. The comparison between calculated Raman active mode frequencies, and available experimental data is given in Tab.~\ref{tab:table8} where a reasonable agreement is observed. In Raman measurements of Ref \cite{Salje1975}, it has been mentioned that the crystal samples were so small that the Raman tensors could not be determined except for the `c' element corresponding to $A_g$ symmetry and these modes are in line with the assignments from our calculations. The peak located at 393 cm$^{-1}$ has only been experimentally measured by Garcia-Sanchez \textit{et al.} \cite{Raul2013} at a frequency of 376 cm$^{-1}$. The reason that this peak is absent in most of the other experimental studies is that it corresponds to irreducible representation $B_g$ and the Raman polarizability tensor for $B_g$ has only non-zero off-diagonal components (see next Section), resulting in a very low Raman intensity (the concentration of the active species is very low in off-diagonal directions)~\cite{Chen2011,Zobeiri2019,Berg2006,Hoang2010,Maczka2009}.

\begin{figure*}

\includegraphics[width=0.86\textwidth]{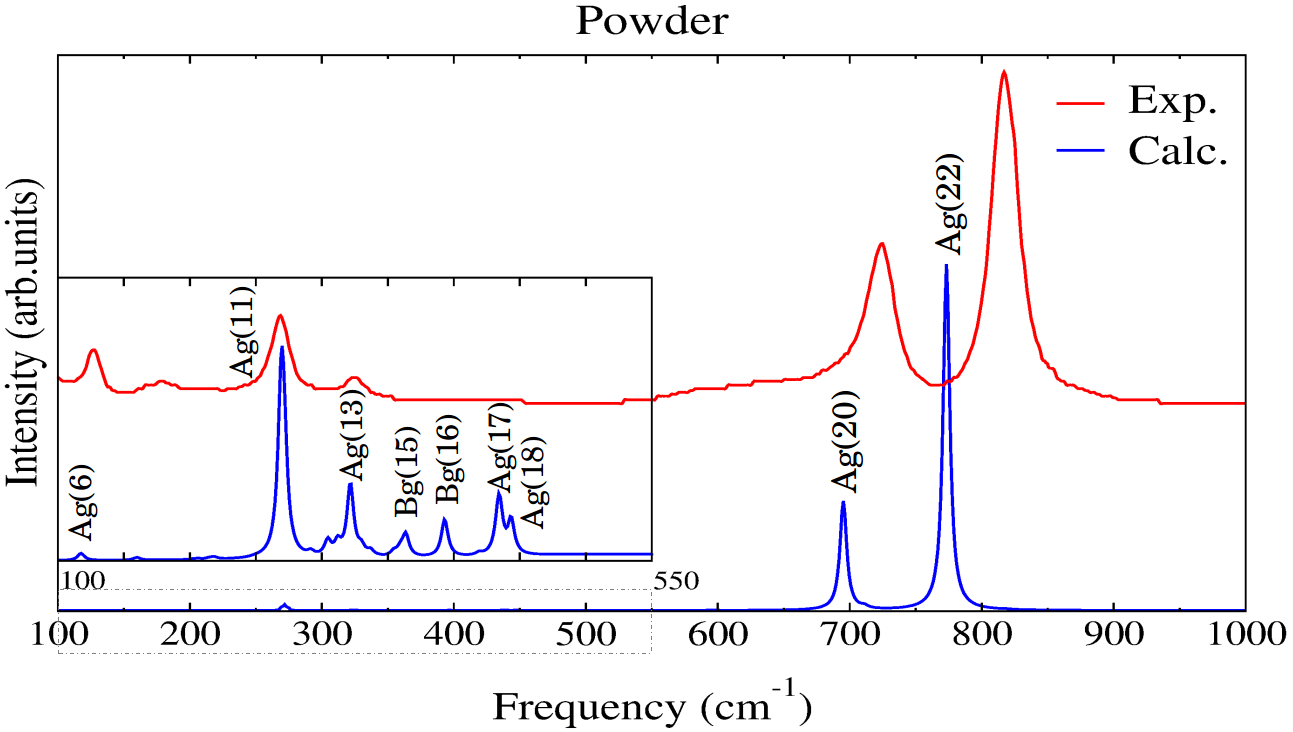}

\caption{\label{fig:fig8}The unpolarized Raman spectra of the $P2_1/n$ phase, the room temperature structure of WO$_3$, the calculated within the LDA (for the visualization purposes the low frequency range (between 100-550 cm$^{-1}$) are magnified), and the experimental Raman spectrum measured on nanoparticle powder of WO$_3$. The spectrum has been acquired using a 532 nm laser at room temperature \cite{Thummavichai2018}.}
\end{figure*}

{\it Raman spectra --} Based on group theory, the Raman susceptibility tensors for the $A_g$ and the $B_g$ modes are given by:
 $A_g$= $\begin{pmatrix}
  a & d & 0\\ 
  d & b & 0\\ 
  0 & 0 & c
  \end{pmatrix}$   ,        $B_g$= $\begin{pmatrix}
  0 & 0 & e\\ 
  0 & 0 & f\\ 
  e & f & 0
  \end{pmatrix} $
so that tensor elements could be determined by the polarized Raman scattering measurement on single crystals in different experimental configurations. As illustrated in Fig.~\ref{fig:fig7}, the calculated polarized Raman spectra show different intensities depending on the geometrical orientation. Unfortunately, no experimental polarized Raman spectra have been reported so far for WO$_3$ single crystals so that our calculations can be used as benchmark results for the interpretation of future measurements. We also determined the powder spectrum by averaging over different orientations of the polarized spectrum. The experimental Raman spectrum measured on nanoparticle powder of WO$_3$ reported in a recent work by Thummavichai \textit{et al.} \cite{Thummavichai2018}, and the calculated powder Raman spectrum are compared in Fig. \ref{fig:fig8}. It appears there that our calculations provide an overall good description for the powder Raman spectrum. In the high-frequency range, the spectrum has two peaks at the correct frequency position and with a relatively good prediction of the intensities. In the low-frequency range peaks are at the right frequency position, but the intensities are strongly underestimated.  

 The two intense peaks in the high-frequency range are dominated by antipolar motions of the oxygen atoms (heavy W atoms are not moving significantly in this frequency range) along $z$ and $y$ orientations (see \ref{sec:level3}). The $A_g$(22) mode at 766 cm$^{-1}$ with the maximum intensity is dominated by the similar antipolar mode at M in the reference cubic phase (antipolar M$^-_3$ mode related to oxygen motions along $z$). The second high-intensity peak at 
689 cm$^{-1}$ is associated to the $A_g$(20) mode which is dominated by the similar antipolar mode at X in the reference cubic phase (antipolar X$^-_5$ mode related to oxygen motions along $y$). Also, in the low-frequency range, we observe peaks associated to modes with dominant antipolar motions, but involving this time both W and O atoms ($A_g$(6) and $A_g$(11)). Other peaks are mostly associated with W-O-W bending modes of the bridging oxygen.

\subsubsection{Phonon dispersion curves}

\begin{figure*}
\hspace*{-1.3cm}
\includegraphics[width=0.92\textwidth]{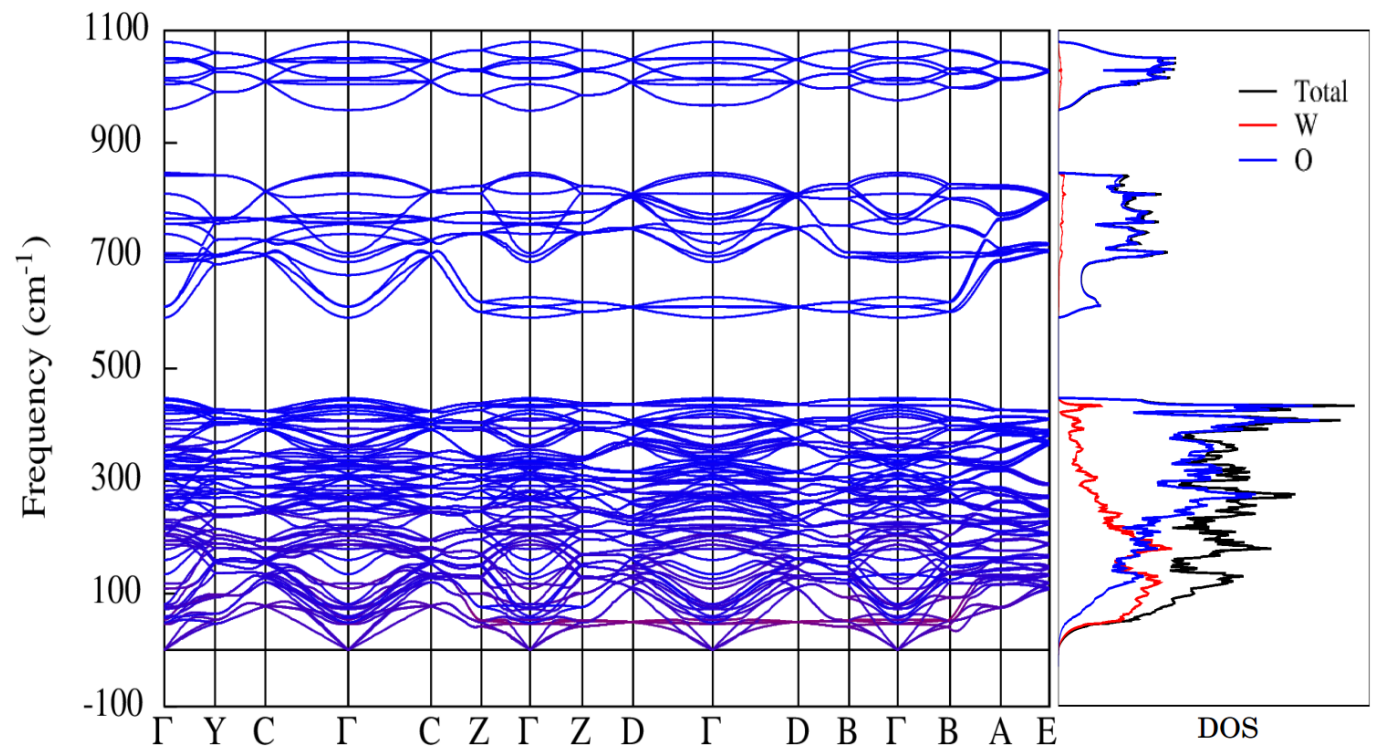}
\caption{\label{fig:fig9}The calculated full phonon dispersion curve and phonon density of states of the room temperature structure $P21/n$ of WO$_3$ within LDA. The blue and red colors distinguish the contributions of oxygen and tungsten atoms, respectively.}
\end{figure*}

Fig.~\ref{fig:fig9} illustrates the phonon dispersion curves and the phonon density of states of the $P2_1/n$ phase. Red and blue colors distinguish the involvement of W and O atoms. There are no imaginary frequencies in the phonon bands, indicating that this phase, although not the ground state, is dynamically stable. The phonon bands are spread into three regions separated by two gaps of 100-150 cm$^{-1}$. The low-frequency region is associated to modes involving motions of W and O atoms while the medium-frequency and high-frequency regions concern nearly pure oxygen motions.   

\subsection{The ground state $P2_1/c$ phase}

In this section, we report the dynamical properties of the $P2_1/c$ phase with the LDA functional and further assess the dynamical stability of the $P2_1/c$ phase with respect to a potential $Pc$ ground state (discussed in Sec.~\ref{sec:level2}). 
\subsubsection{Irreducible representations at $\Gamma$}
The non-polar $P2_1/c$ structure is similar to the $P2_1/n$ phase and it belongs to the $C^5_{2h}$ point group. However, the primitive unit cell is rotated by 45$^{\circ}$ and only contains 16 atoms so that there are 48 $\Gamma$-phonon modes at each $k$-point. The zone-center phonons can be classified according to the irreducible representations of this point group as $$\Gamma_{phonons} = 12 A_g \oplus 12 A_u \oplus 12 B_g \oplus 12 B_u .$$ These include: (i)  three acoustic modes ($\Gamma_{acoustic}= A_{u}+ 2B_{u}$), (ii) 24 Raman active modes ($A_g$ and $B_g$) and (iii) 21 infrared (IR) active modes ($A_u$ and $B_u$). 
\subsubsection{Infrared active modes}
{\it TO modes --} In the first part of Tab.~\ref{tab:table9} we report the calculated TO frequencies of the infrared active modes. According to this Table, there is no unstable mode at $\Gamma$: the lowest polar mode has a frequency of 139 cm$^{-1}$ so that it is far from being unstable and the result appears robust (in agreement with Hamdi et al. \cite{Hamdi2016}). 

{\it LO modes --} The investigation of the LO modes along the different high-symmetry directions of the Brillouin zone, their corresponding irreducible representations, and mode effective charges $\overline{Z}^*$ of the infrared active modes along the $x$, $y$ and $z$ direction are presented in Tab.~\ref{tab:s4} of the appendix. This Table shows that along all the high-symmetry directions of the Brillouin zone, there are only three large LO-TO splittings in the same LO frequency ranges: at 260 $\pm$ 20 cm$^{-1}$, 415 $\pm$ 10 cm$^{-1}$, and 960 $\pm$ 20 cm$^{-1}$. In addition, the strongest LO-TO splitting is located at 962 cm$^{-1}$ with a width of 322 cm$^{-1}$ and is associated to the $B_u$(8) mode in the $[1 0 1]$ direction. This mode is associated with the opposite motions of W and O along the $x$ direction. 

{\it IR reflectivity spectra --} The infrared reflectivity spectra are also calculated at normal incidence for all the Cartesian directions. Comparing to the reflectivity spectra of the $P2_1/n$ phase, here, between 750 to 800 cm$^{-1}$ two peaks of $A_u$ and $B_u(z)$ are absent. This point can be exploited for distinguishing these two phases in future infrared measurements. 
\begin{figure} 
\hspace*{-0.8cm}
\includegraphics[width=0.5\textwidth]{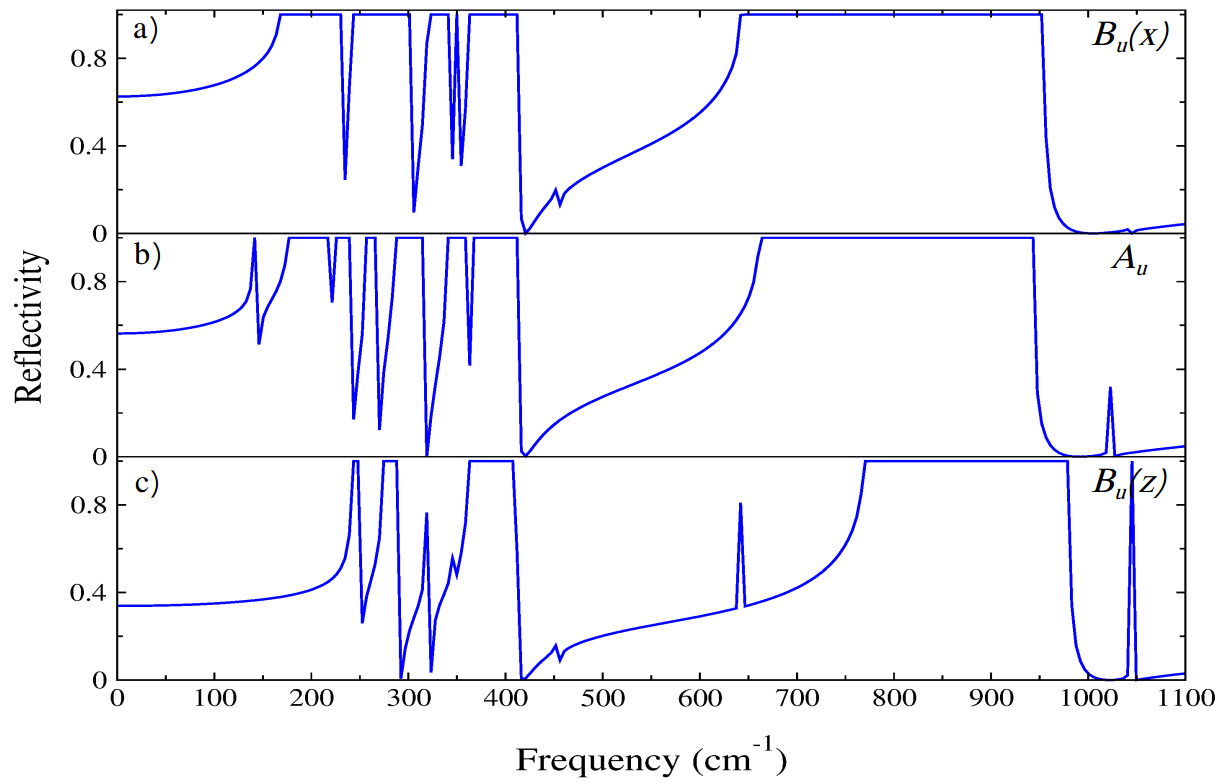}

\caption{\label{fig:fig10}The calculated infrared reflectivity of the $P2_1/c$ phase: (a) $B_u$ with the polarity along $x$, (b) $A_u$, (c) $B_u$ with the polarity along $z$.}
\end{figure}

\begin{figure}
\hspace*{-0.8cm}
\includegraphics[width=0.53\textwidth]{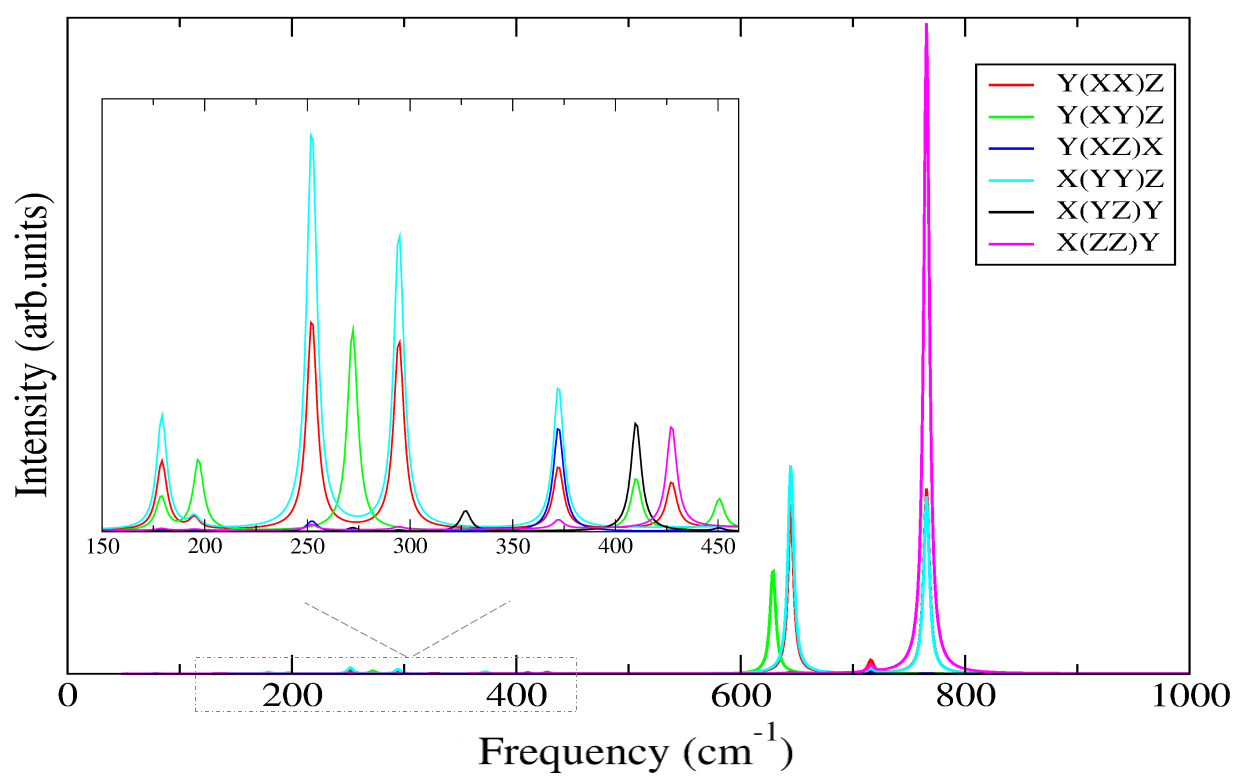} 
\caption{\label{fig:fig11}The polarized Raman spectra of the $P2_1/c$ phase of WO$_3$ calculated within the LDA (for visualization purposes the low frequency range between 450-150 cm$^{-1}$ is magnified).}
\end{figure}

\begin{figure*} 
\includegraphics[width=0.86\textwidth]{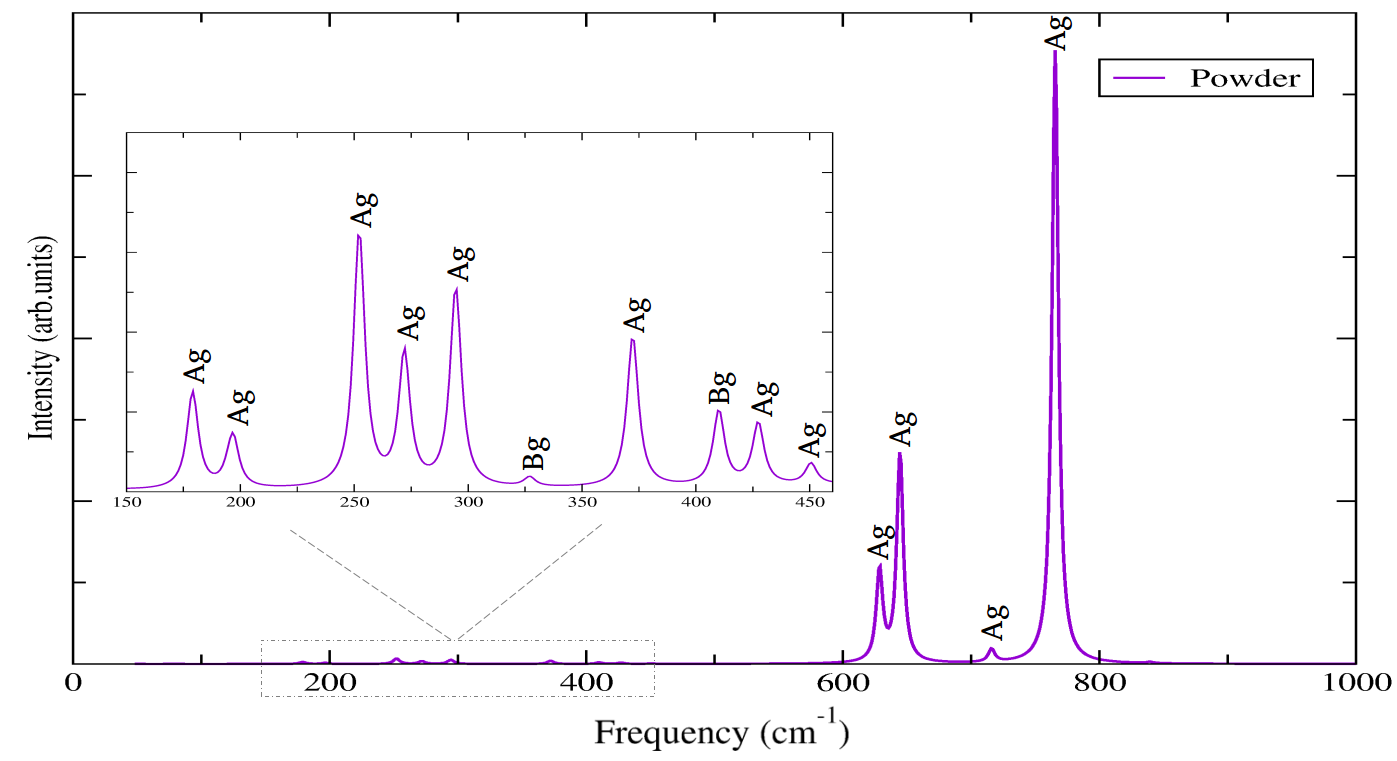} 
\caption{\label{fig:fig12}The unpolarized Raman spectra of the $P2_1/c$ phase, the ground state of WO$_3$, calculated within LDA (for the visualization purposes the low frequency range (between 450-150 cm$^{-1}$) is magnified).}
\end{figure*}

\begin{table}
\caption{\label{tab:table9}The TO phonon frequencies of the ground state $P2_1/c$ phase at $\Gamma$ within LDA.} 
\begin{ruledtabular}
\begin{tabular}{cccc}
\textrm{Irrep}&
\multicolumn{1}{c}{\textrm{Freq(cm$^{-1}$)}}&
\multicolumn{1}{c}{\textrm{Irrep}}& \textrm{Freq(cm$^{-1}$)} \\
\colrule
$A_u$ &139 &$B_u$ &166\\
$A_u$ &174&$B_u$&241\\
$A_u$ &221&$B_u$&272\\
$A_u$ &255&$B_u$ &318\\
$A_u$ &284&$B_u$&345\\
$A_u$ &302&$B_u$&360\\
$A_u$ &339&$B_u$&453\\
$A_u$ &365&$B_u$&640\\
$A_u$ &659&$B_u$&766\\
$A_u$ &737&$B_u$&1041\\
$A_u$ &1020&&\\
\cline{1-4}
$A_g$ & 50&$B_g$&53\\
$A_g$ & 79 &$B_g$&76\\
$A_g$ & 136&$B_g$ &178\\
$A_g$ &179&$B_g$&196 \\
$A_g$ &195&$B_g$& 272\\
$A_g$ &252& $B_g$ & 327\\
$A_g$ &294&$B_g$&410\\
$A_g$ &372&$B_g$&423\\
$A_g$&427&$B_g$ &450\\
$A_g$&644&$B_g$& 628\\
$A_g$ &716&$B_g$&839\\
$A_g$ &765&$B_g$ &1076\\
\end{tabular}
\end{ruledtabular}
\end{table}
\begin{figure*}
\hspace*{-1.3cm}
\includegraphics[width=0.92\textwidth]{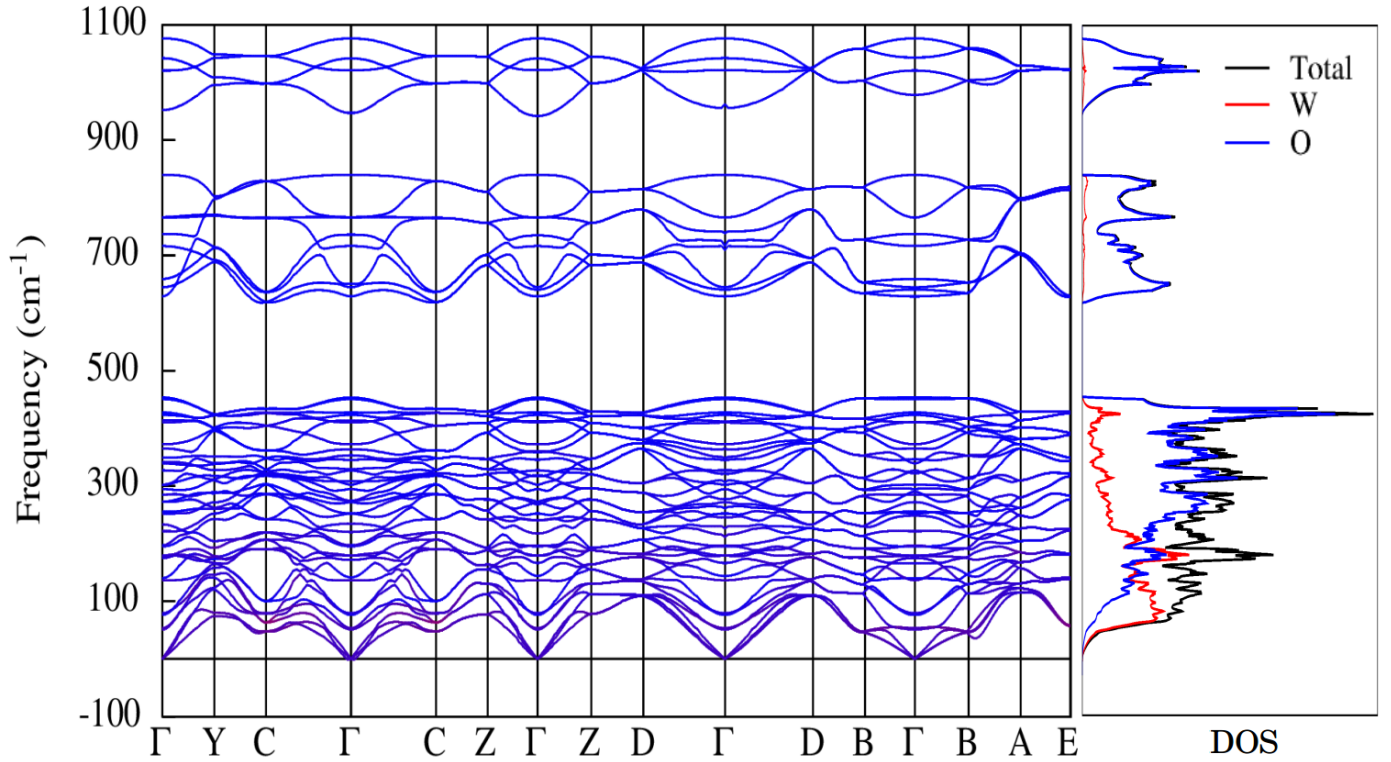}
\caption{\label{fig:fig13}The calculated full phonon dispersion curves and phonon density of states of the ground state $P2_1/c$ phase of WO$_3$ within the LDA. The blue and red colors represent the contribution of oxygen and tungsten atoms, respectively.}
\end{figure*}

\subsubsection{Raman active modes}
{\it Raman modes --} In the second part of Tab.~\ref{tab:table9} we report the calculated TO frequencies of the Raman active modes classified according to their irreducible representations. We are not aware of any experimental data nor other theoretical values for comparison.

{\it Raman spectra --} The calculated polarized, and unpolarized Raman spectra of the $P2_1/c$ phase are presented in Fig.~\ref{fig:fig11} and Fig.~\ref{fig:fig12}, respectively. The Raman susceptibility tensors for the $A_g$ and $B_g$ modes are the same as in the $P2_1/n$ phase. Although the number of Raman active modes of the $P2_1/c$ phase are half the number of Raman active modes of the $P2_1/n$ phase, more peaks are observed in the Raman spectrum of the $P2_1/c$ phase, because the Raman intensities are higher in the $P2_1/c$ phase. This might be because in this phase the density of identical bond characters are more pronounced. 
In the high-frequency range, the powder Raman spectrum of the $P2_1/n$ phase shows only two intense peaks, whereas there are three intense and one small peak in $P2_1/c$ phase spectrum. The small peak at 716 cm$^{-1}$ corresponds mainly to the Jahn–Teller distortion. The mode with the highest intensity at 766 cm$^{-1}$ is dominated by the antipolar oxygen motions along the $z$ direction (related to M$^-_3$ from the reference cubic phase, see \ref{sec:level3}). Other two peaks with noticeable intensities are also dominated by the antipolar oxygen motions but this time in the $xy$-plane (X$^-_5$ antipolar mode from the reference cubic phase). 

As a result, the observation of the high-frequency range of the spectrum can provide a clear way to distinguish the $P2_1/n$ and $P2_1/c$ phases. More generally, it is worth noticing that the number of intense peaks in the high-frequency range reflects the number of antipolar distortions that have been condensed in the crystal structure. Indeed, condensing an anti-polar distortion transforms the related anti-polar mode originally at the zone boundary into a Raman active mode at $\Gamma$. In the $P2_1/n$ (resp. $P2_1/c$) phase, we have antipolar distortions along $z$ and $y$ directions (resp. $z$, $y$ and $x$ directions) and two (resp. three) intense peaks. In order to further validate this statement, we calculated also the Raman spectrum of $P4/ncc$ and $P4/nmm$ phases, both containing only one antipolar distortion along $z$ (M$^-_3$) and, in both cases, only one intense peak appears in the high-frequency range.

Comparing the Raman spectrum of the $P2_1/n$ and $P2_1/c$ phases, we further observe a small shift of the peaks. The larger the alternation of long-short $O-W-O$ bonds, the higher is the frequency of the peaks. The 766 cm$^{-1}$ peak is an exact match. This is expected because $P2_1/c$ and $P2_1/n$ phases have the same long-short $O-W-O$ bonds length equal to 1.76-2.09 \r{A} along $z$. The 689 cm$^{-1}$ peak in the Raman spectrum of the $P2_1/n$ is attributed to the long-short $O-W-O$ bonds equal to 1.78-2.04 \r{A} along $y$, while in the $P2_1/c$ phase, the second peak is shifted to the lower frequency (644 cm$^{-1}$) due to the smaller alternation of long-short $O-W-O$ bonds (1.81-1.95 \r{A} along $y$).

\subsubsection{Phonon dispersion curves}

The phonon dispersion curves and phonon projected density of states of the $P2_1/c$ phase are shown in Fig.~\ref{fig:fig13}. Similarly to the $P2_1/n$ phase, there is no imaginary frequency, confirming the dynamical stability of this phase. Again, the frequencies are spread into 3 well-separated regions. The main differences between the two phases appear in the intermediate region, involving the stretching vibration of O-W-O. Also the DOS of the low-energy region is shifted to slightly higher frequencies in the $P2_1/c$ phase.

\section{Conclusions}
In this work we have verified the ability of various DFT functionals (B1-WC, HSE06, LDA, GGA-WC, PBEsol) to describe the structural energetics of all the experimentally observed phases as well as six additional metastable phases of WO$_3$. We have analyzed the atomic distortions of these phases in terms of symmetry adapted modes of the hypothetical cubic phase and quantified the amplitude of the distortions. We have analyzed their energetics based on the instabilities presented in the phonon dispersion curve of the parent cubic phase. It can be concluded that (i) the B1-WC hybrid functional is the most appropriate to describe WO$_3$ while (ii)  among the investigated standard functionals, LDA is the most suitable to reproduce the structural distortions of the different phases of WO$_3$ and to predict the correct ground state. We have also re-discussed briefly the unusual phase diagram of WO$_3$.

We have exhaustively described the dielectric and lattice dynamical properties of the ground state $P2_1c$ phase and room temperature $P2_1/n$ phase. The Born effective charges, dielectric tensors and the refractive indices were obtained. The phonon dispersion curves were also reported; no imaginary frequency was observed, indicating that both phases are dynamically stable. The $\Gamma$-phonon modes of both phases were carefully studied and compared to available data. The infrared reflectivity and Raman spectra were also calculated and reasonable agreement between experimental and calculated spectra was found for the room temperature phase. Also, we have carefully discussed the Raman spectra, explaining the physical origin of their main features and evolution from one phase to another. We pointed out that the alternation of long-short bonds along $O-W-O$ chains can lead to important consequences on the dynamical properties of different phases of WO$_3$. 
We also revealed that the number of peaks in the high-frequency range of the Raman spectrum appears as a fingerprint of the number of antipolar distortions that are present in the structure so that measurement of the Raman spectrum of WO$_3$ appears as an efficient way to distinguish between its different phases. 

Our calculations appear as benchmark results for the interpretation of future experimental measurements. Furthermore, many physical properties of WO$_3$ are affected by polarons, which arise from the coupling of excess charge carriers with specific phonon modes. Our work provides all the ingredients for future more systematic analysis of which phonon modes are involved in polaron formation and key to localize the charge. 


\section*{Acknowledgements}
This work has been funded by the Communaut\'e Fran\c{c}aise de Belgique (ARC AIMED G.A. 15/19-09) and a Methusalem project of the University of Antwerp .
EB thanks the FRS-FNRS for support. 
The authors acknowledge the CECI supercomputer facilities funded by the F.R.S-FNRS (Grant No. 2.5020.1), the Tier-1 supercomputer of the F\'ed\'eration Wallonie-Bruxelles funded by the Walloon Region (Grant No. 1117545), and the computing facilities of the Flemish Supercomputer Center.
We acknowledge that the results of this research have been achieved using the DECI resource BEM based in Poland at Wrocław with support from the PRACE OFFSPRING project.


\appendix

\setcounter{equation}{0}
\setcounter{figure}{0}
\setcounter{table}{0}
\setcounter{section}{0}
\makeatletter
\renewcommand{\theequation}{A\arabic{equation}}
\renewcommand{\thetable}{A\arabic{table}}
\renewcommand{\thefigure}{A\arabic{figure}}
\renewcommand{\citenumfont}[1]{A#1}
\section{Phonon dispersion calculations}

\begin{figure}[h]
\begin{subfigure}{0.45\textwidth}
 \hspace*{0.0cm} 
\includegraphics[width=\textwidth,height=5cm]{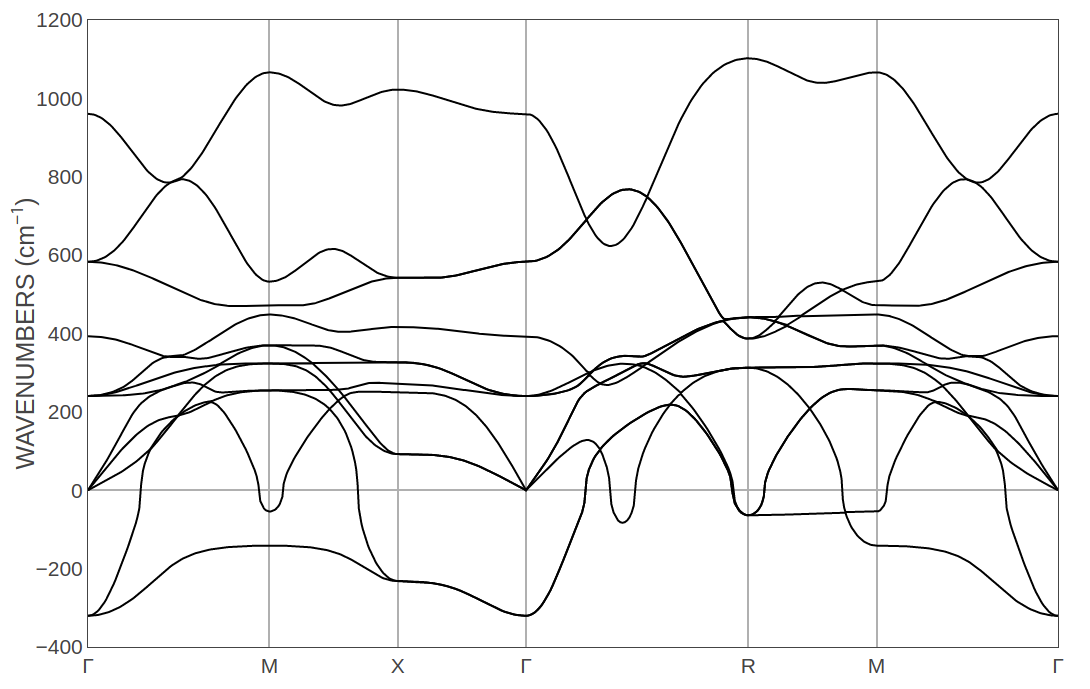} 
 \caption{}\label{fig:A1a}
\end{subfigure}
\begin{subfigure}{0.49\textwidth} \hspace*{-1.0cm}
\includegraphics[width=\textwidth]{fig1.png} 
\caption{}\label{fig:A1b}
\end{subfigure}
\caption{The phonon dispersion curves of the cubic phase of WO$_3$, obtained by (a) the CRYSTAL17 code, and the improper implementation of the mixed-space approach, and (b) the ABINIT code, and the proper employment of Fourier interpolation technique.}
\end{figure}

In the CRYSTAL17 package, the phonon frequencies are computed using the frozen phonon numerical differences approach. Fig.~\ref{fig:A1a} shows the phonon bands of the parent cubic phase of WO$_3$ generated by CRYSTAL17. The unphysical oscillations and intersections observed in the high frequency range of the phonon bands are due to the improper implementation of the mixed-space approach \cite{Wang2016} (the cancellation of the artificial macroscopic electric field is not applied) which causes problems in Fourier interpolations at arbitrary, non high symmetry q-points. As a result, the correct phonon bands has been obtained just after a proper use of the Fourier interpolation technique (shown in Fig.~\ref{fig:A1b}).
\begin{figure}\label{fig:A2}
\begin{subfigure}{0.44\textwidth}
 \hspace*{0.0cm} 
\includegraphics[width=\textwidth]{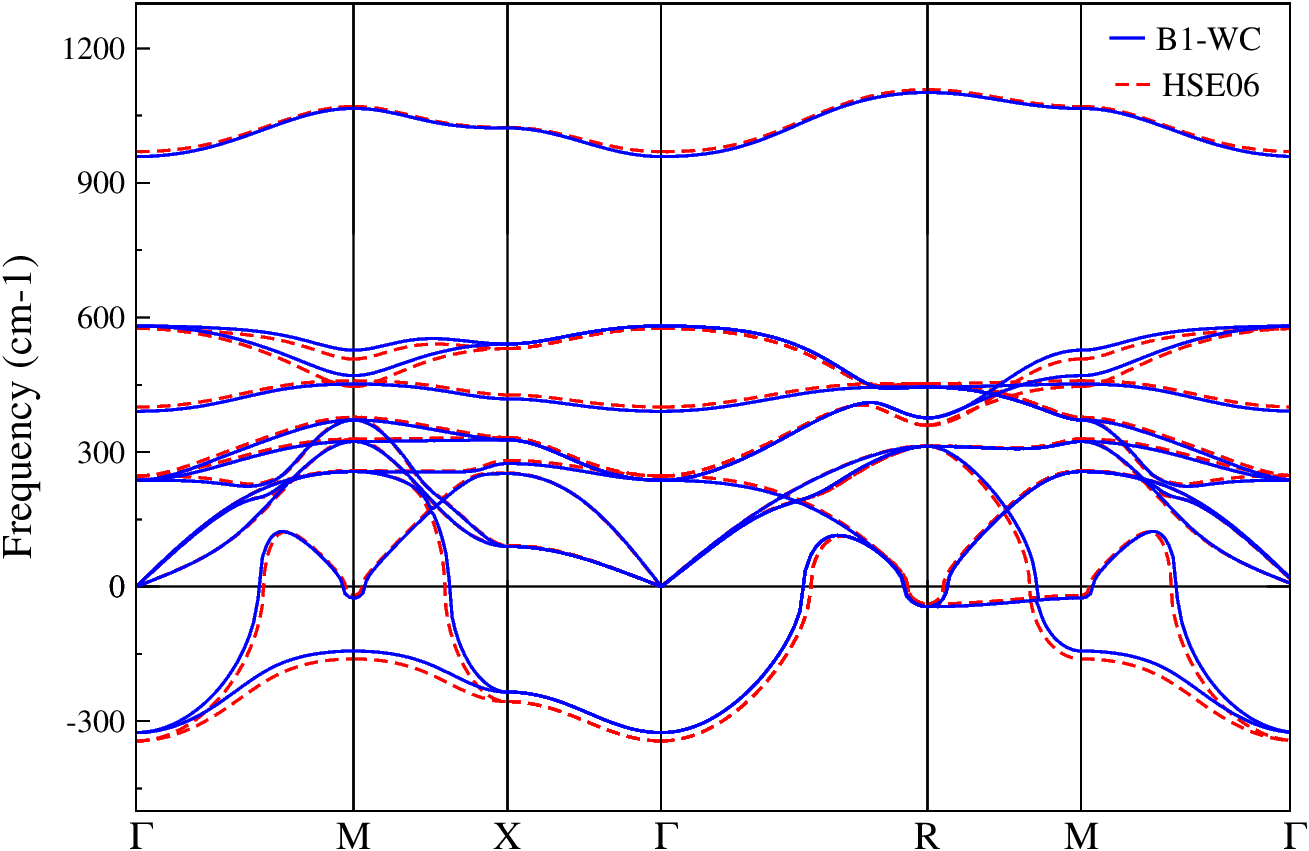} 
 \caption{}\label{fig:A2-1}
\end{subfigure}
\begin{subfigure}{0.44\textwidth}\hspace*{0.0cm}
\includegraphics[width=\textwidth]{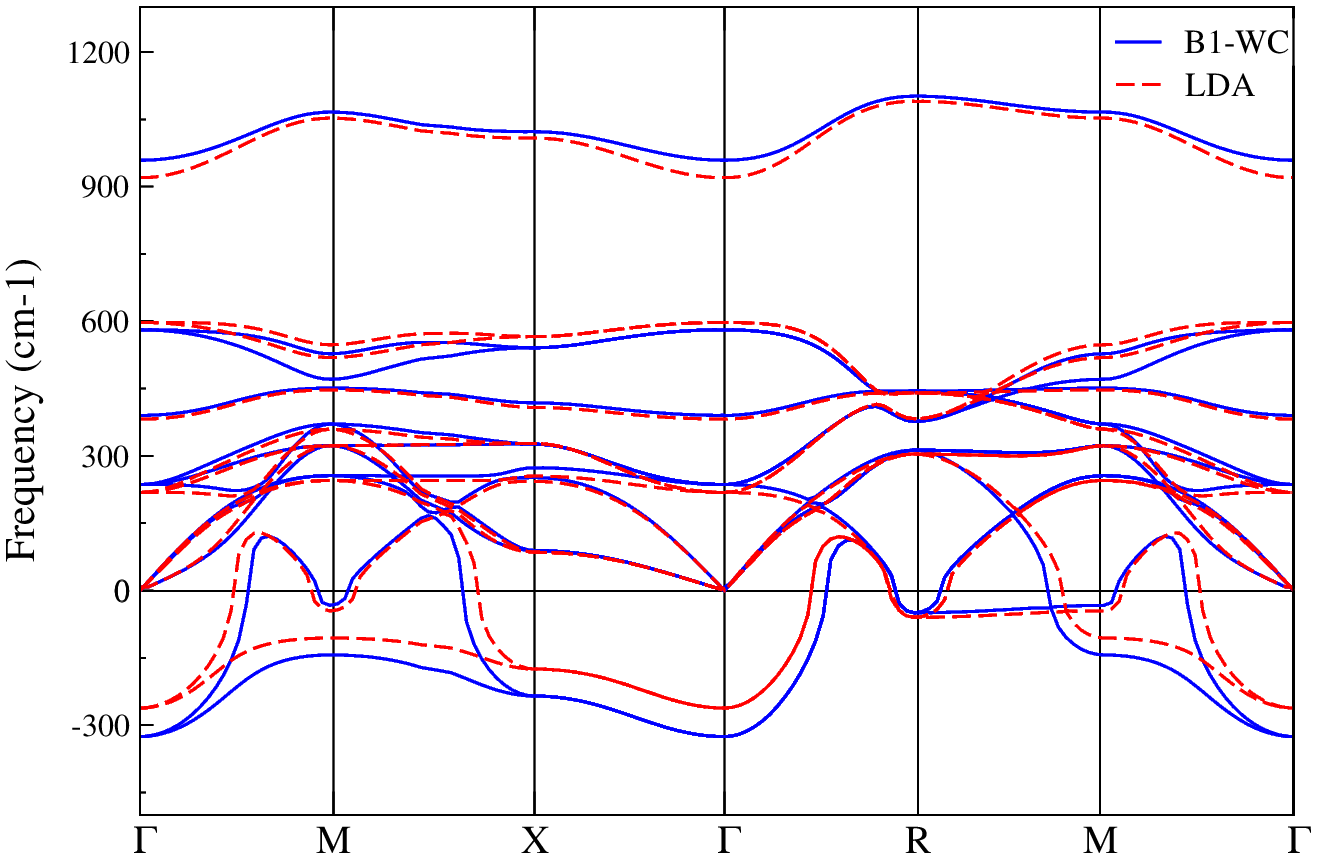} 
\caption{}\label{fig:A2-2}
\end{subfigure}
\begin{subfigure}{0.44\textwidth}
 \hspace*{0.0cm} 
\includegraphics[width=\textwidth]{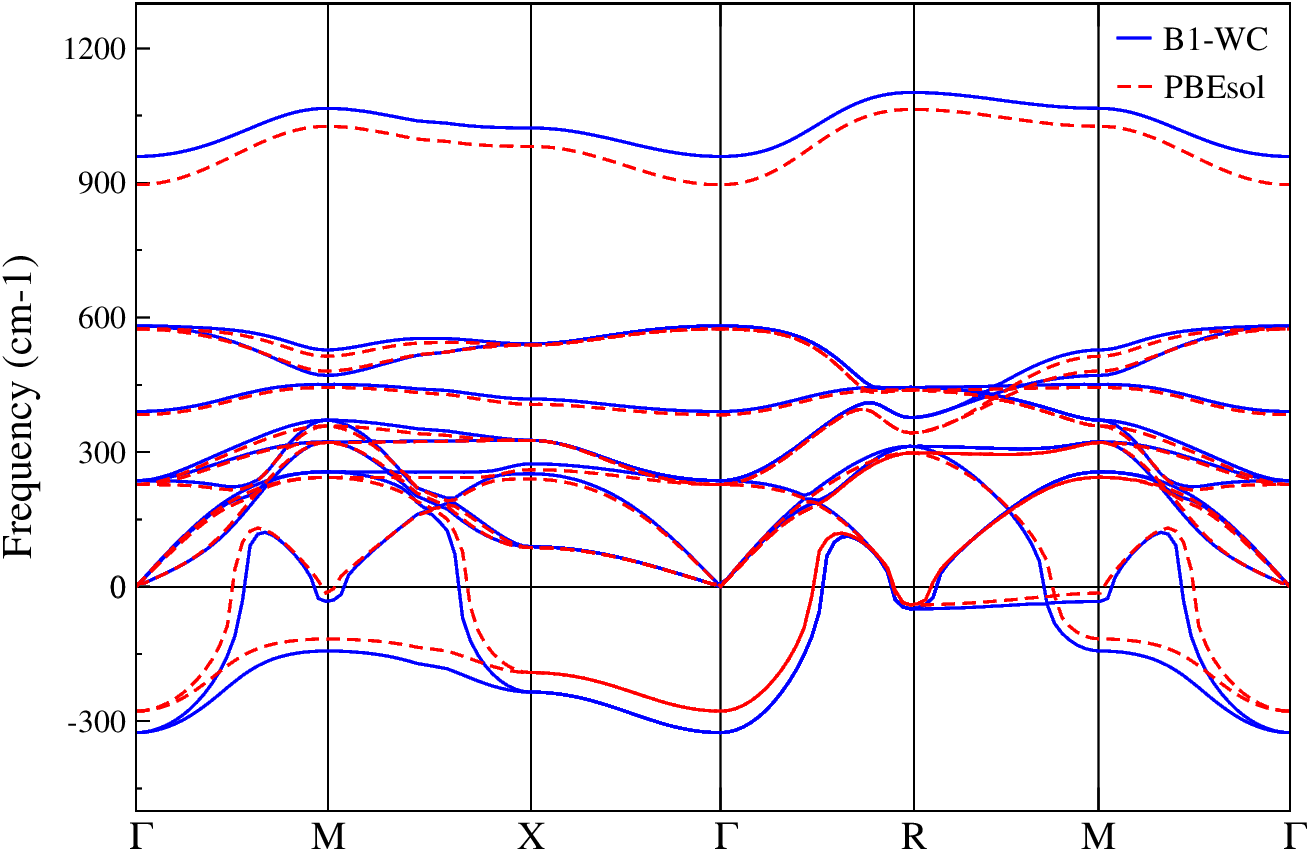} 
 \caption{}\label{fig:A2-3}
\end{subfigure}
\begin{subfigure}{0.44\textwidth}\hspace*{0.0cm}
\includegraphics[width=\textwidth]{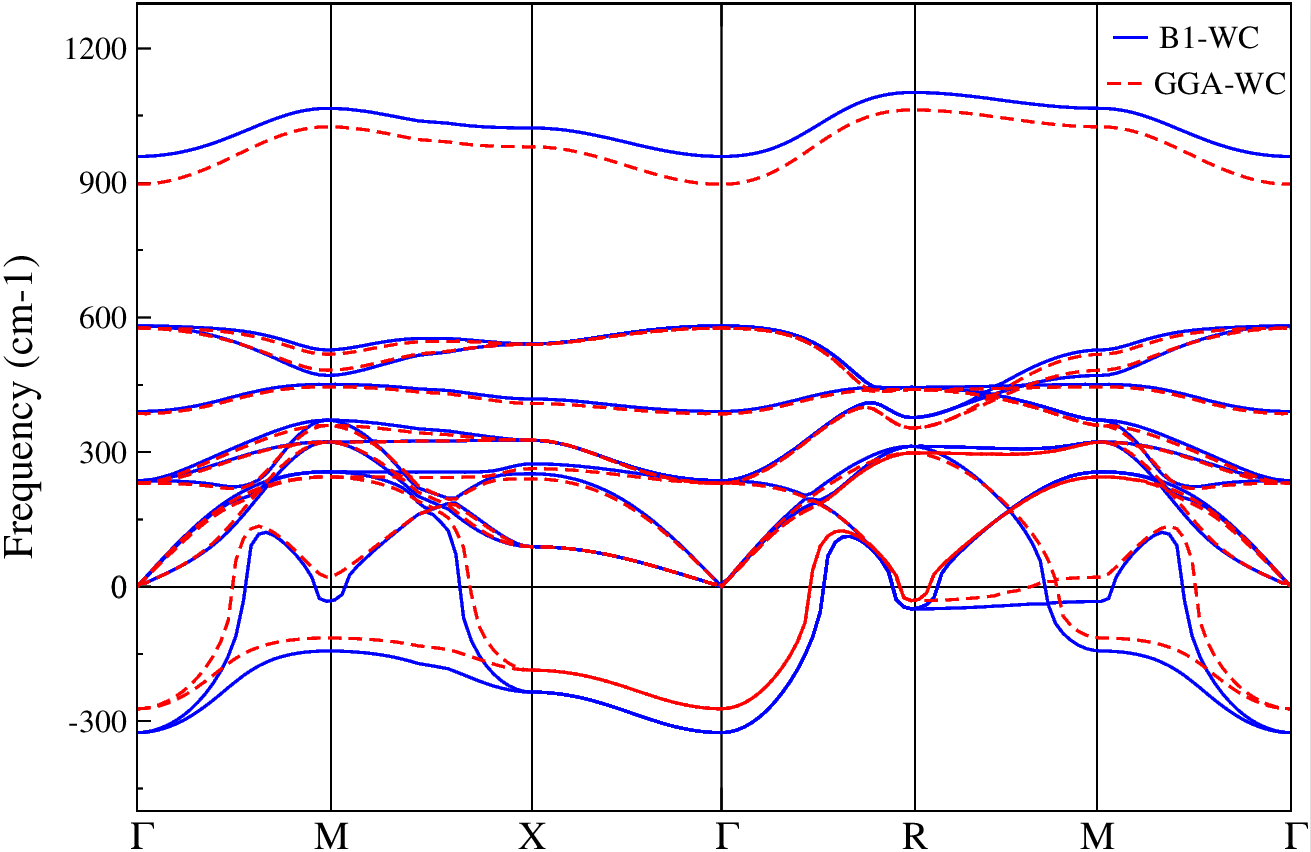} 
\caption{}\label{fig:A2-4}
\end{subfigure}
\caption{The comparison of phonon dispersion curves of the cubic phase obtained by B1-WC and (a) HSE06, (b) LDA, (c) PBEsol, (d) GGA-WC.}\label{fig:s3}
\end{figure}
\begin{table}
\caption{\label{tab:tableS44}%
Atomic positions of $P2_1/n$ and $P2_1/c$ phases of $WO_3$ in reduced coordinates. Only atomic positions are relaxed by LDA with the fixed lattice constants of the experimental values, and calculated within B1-WC  for $P2_1/n$ (7.30, 7.53, 7.69 \AA), and $P2_1/c$ (5.26, 5.15, 7.61 \AA), respectively.}

\begin{ruledtabular}
\begin{tabular}{lccc}
\textrm{Phase}&
\multicolumn{1}{l}{\textrm{atom}}&
\multicolumn{1}{l}{\textrm{Wyckoff}}&
\multicolumn{1}{c}{\textrm{Position}}\\

\colrule

$P2_1/n$& $W_1$  & 4e & 0.2487 0.0274  0.2203  \\
& $W_2$   & 4e & 0.2507  0.0286  0.7211\\
& $O_1$   & 4e &  0.5011 0.0338  0.2713 \\
& $O_2$   & 4e &  0.4993  0.4657  0.2707  \\
& $O_3$   & 4e & 0.2158 0.2611  0.2380 \\
& $O_4$   & 4e &    0.2851 0.2611 0.7509 \\
& $O_5$   & 4e &  0.2269 0.0199  0.4927 \\
& $O_6$   & 4e & 0.2281 0.5067  0.5067 \\
\colrule
$P2_1/c$& $W_1$  & 4e & 0.2541 -0.2338 -0.2180  \\
& $O_1$   & 4e & 0.0343  0.0383 -0.2209 \\
& $O_2$   & 4e & 0.4580 0.4637 -0.2794  \\
& $O_3$   & 4e &  0.2521 -0.3077  0.0078 \\

\end{tabular}
\end{ruledtabular}
\end{table}
\begin{table}
\caption{\label{tab:s1}
The Born effective charge tensors ($Z^*$) of the $P2_1/c$ phase of WO$_3$, calculated by both LDA and B1-WC approaches (see notations of Tab.~\ref{tab:tableS44}).
}
\begin{ruledtabular}
\begin{tabular}{cc}
\textrm{B1-WC}&
\textrm{LDA}\\
\colrule
$Z^*$($W_1$)   & $Z^*$($W_1$)  \\
   $\begin{pmatrix}
  9.97 & -1.07 & -0.16\\ 
  0.9 & 8.78 & 1.05\\ 
  -0.19 & -1.09 & 7.44
\end{pmatrix}$                      &  $\begin{pmatrix}
  10.38 & -1.07 & -0.15\\ 
  0.99 & 9.40 & 0.95\\ 
  -0.17 & -0.89 & 7.74
\end{pmatrix}$                                 \\ 
\colrule
$Z^*$($O_1$)  & $Z^*$($O_1$) \\
    $\begin{pmatrix}
  -4.4 & 3.04 & -0.63\\ 
  2.75 & -3.87 & 0.47\\ 
  -0.28 & 0.26 & 1
\end{pmatrix}$                          &     $\begin{pmatrix}
  -4.57 & 3.05 & -0.66\\ 
  2.92 & -4.13 & 0.54\\ 
  -0.24 & 0.23 & -1.06
\end{pmatrix}$                          \\

\colrule
$Z^*$($O_2$)  & $Z^*$($O_2$) \\
  $\begin{pmatrix}
  -4.52 & 2.83 & 0.6\\ 
  2.71 & -3.73 & -0.54\\ 
  0.29 & 0.3 & -1.04
\end{pmatrix}$                       &  $\begin{pmatrix}
  -4.62 & 2.90 & 0.65\\ 
  2.89 & -4.01 & -0.57\\ 
  0.23 & -0.23 & -1.11
\end{pmatrix}$                        \\

\colrule
$Z^*$($O_3$)   & $Z^*$($O_3$) \\
   $\begin{pmatrix}
  -1.05 & -0.006 & 0.19\\ 
  -0.04 & -1.2 & 0.34\\ 
  0.17 & 0.6 & -5.42
\end{pmatrix}$                            &$\begin{pmatrix}
  -1.19 & 0.00 & 0.16\\ 
  -0.03 & -1.25 & 0.21\\ 
  0.18 & 0.45 & -5.56
\end{pmatrix}$                                \\

\end{tabular}
\end{ruledtabular}
\end{table}
\begin{table}
\caption{\label{tab:s2}%
Comparison between optical dielectric tensor (${\varepsilon }^{\infty }_{ij}$) of the $P2_1/c$ phase of WO$_3$, calculated by HSE06, B1-WC hybrid functional and LDA.
}
\begin{ruledtabular}
\begin{tabular}{ccc}
\textrm{HSE06}&
\textrm{B1-WC}&
\textrm{LDA}\\
\colrule

$\begin{pmatrix}
 5.34&0&-0.09\\ 
  0&4.94&0\\ 
  -0.09&0&4.40
\end{pmatrix}$ &
   $\begin{pmatrix}
  6.06 & 0 & -0.09\\ 
  0 & 5.53 & 0\\ 
  -0.09 & 0 & 4.87
\end{pmatrix}$ 
&  $\begin{pmatrix}
  6.95 & 0 & -0.08\\ 
  0 & 6.50 & 0\\ 
  -0.08 & 0 & 5.44
\end{pmatrix}$  
\end{tabular}
\end{ruledtabular}
\end{table}
\section{Comparison of LDA and B1-WC for lattice dynamical properties of $P2_1/c$ phase}
In order to be able to accurately match the $\Gamma$- phonon modes obtained by LDA and B1-WC, we have evaluated the overlap matrix between the phonon eigenmodes computed from the diagonalization of the dynamical matrix by ABINIT/LDA (${\eta }_A$) and those computed by CRYSTAL/B1-WC (${\eta }_C$), namely $\left\langle {\eta }_C\mathrel{\left|\vphantom{{\eta }_C M {\eta }_A}\right.\kern-\nulldelimiterspace}M\mathrel{\left|\vphantom{{\eta }_C M {\eta }_A}\right.\kern-\nulldelimiterspace}{\eta }_A\right\rangle $, where $M$ is the mass matrix. In case of two corresponding eigenmodes, this overlap $\left\langle {\eta }_C\mathrel{\left|\vphantom{{\eta }_C M {\eta }_A}\right.\kern-\nulldelimiterspace}M\mathrel{\left|\vphantom{{\eta }_C M {\eta }_A}\right.\kern-\nulldelimiterspace}{\eta }_A\right\rangle $ should be approximately 1. We could realize such a one-to-one correspondence between the phonon eigenmodes, and thus also a match between the corresponding phonon frequencies, which are shown in Tab.~\ref{tab:s3}. The comparison of the calculated phonon frequencies by both LDA and B1-WC indicates that they are in good agreement. 
\begin{table} 
\caption{\label{tab:s3}%
The TO phonon frequencies of the ground state $P2_1/c$ phase at $\Gamma$ within the B1-WC and LDA approaches, in cm$^{-1}$, and the overlap matrix between the phonon eigenmodes computed from the diagonalization of the dynamical matrix by ABINIT/LDA (${\eta }_A$) and those computed by CRYSTAL/B1-WC (${\eta }_C$).
}
\begin{ruledtabular}
\begin{tabular}{cccc}
\textrm{Irrep}&
\multicolumn{1}{c}{\textrm{B1-WC}}&
\multicolumn{1}{c}{\textrm{LDA}}& \textrm{$\left\langle {\eta }_C\mathrel{\left|\vphantom{{\eta }_C M {\eta }_A}\right.\kern-\nulldelimiterspace}M\mathrel{\left|\vphantom{{\eta }_C M {\eta }_A}\right.\kern-\nulldelimiterspace}{\eta }_A\right\rangle $} \\
\colrule
$A_g$ & 59 & 50 & 1.00\\
$A_g$ & 93 & 79 &1.00 \\
$A_g$ & 143 & 136& 0.97\\
$A_g$ &183 & 179  & 0.92\\
$A_g$ & 215 &  195 & 0.93\\
$A_g$ & 269 &252  &0.99\\
$A_g$ & 310&294   & 0.98\\
$A_g$ & 386 & 372  & 0.98\\
$A_g$ &429  & 427 &1.00\\
$A_g$ &681 & 644  & 1.00\\
$A_g$ &750  & 716  & 1.00\\
$A_g$ & 805 & 765 & 1.00\\
\cline{1-4}
$B_g$&64 & 53  & 0.98 \\
$B_g$& 91& 76&  0.98\\
$B_g$ & 204 &178 & 0.92\\
$B_g$&174 &196   & 0.91 \\
$B_g$& 281&272 &  0.96\\
$B_g$ &  332&327 & 0.97\\
$B_g$&429&  410 & 0.96\\
$B_g$&418 & 423&  0.99\\
$B_g$ & 457&450 & 0.96\\
$B_g$&640& 628  &1.00 \\
$B_g$&866 & 839&  1.00\\
$B_g$ & 1083 & 1076& 1.00\\
\cline{1-4}
$A_u$ &144  &139 & 1.00\\
$A_u$ & 184 & 174& 0.98\\
$A_u$ & 216 &221 & 0.98\\
$A_u$ & 262& 255& 0.96\\
$A_u$ & 287 & 284& 0.95\\
$A_u$ &308  &302 & 0.96\\
$A_u$ &359  & 339& 0.98\\
$A_u$ & 383 &365 & 0.98\\
$A_u$ & 698 &659 & 1.00\\
$A_u$ &765  &737 & 1.00\\
$A_u$ & 1038 &1020 & 1.00\\
\cline{1-4}
$B_u$ &188 &166&0.99 \\
$B_u$ &234 &241& 1.00\\
$B_u$ &276 &272& 1.00\\
$B_u$ &327 &318& 0.99\\
$B_u$ &350 &345& 0.99\\
$B_u$ &383 &360&0.99 \\
$B_u$ &456 &453& 1.00\\
$B_u$ &651 &640& 1.00\\
$B_u$ &796 &766& 1.00\\
$B_u$ & 1052&1041& 1.00\\
\end{tabular}
\end{ruledtabular}
\end{table}
\section{LO-TO modes}
In Tab.~\ref{tab:s4} and Tab.~\ref{tab:s5}, the LO frequencies of the IR active modes for the $P2_1/c$ and $P2_1/n$ phase, respectively, are reported along different directions together with related TO frequencies and mode effective charges.
\begin{table*}
\caption{\label{tab:s4} The LO frequencies (in cm$^{-1}$) of the infrared active modes along the different high-symmetry directions of the Brillouin zone, their corresponding irreducible representations, and mode effective charges ($\overline{Z}^*$) of the TO infrared active modes along the $x$, $y$ and $z$ is presented for the $P2_1/c$ phase calculated by LDA. The frequencies shown by light gray in the LO columns are equal to the TO frequencies. The frequencies shown in bold font present the strong shifts in LO-TO splitting. The coordinates of the high-symmetry points are as follows: Y(1/2,0,0), Z(0,1/2,0), B(0,0,1/2), C(1/2,1/2,0), D(0,1/2,1/2), A(1/2,0,1/2) and E(1/2,1/2,1/2).}
\begin{ruledtabular}
\begin{tabular}{cccccccccccc}
\textbf{Irrep} &\textbf{TO Modes} & \multicolumn{3}{c}{\textbf{$\overline{Z}^*$}}  & \multicolumn{7}{c}{\textbf{LO Modes }} \\
\cline{3-5} \cline{6-12}
& &\textbf {x} &\textbf{y} &\textbf{z} &\textbf{Y} & \textbf {Z}  &\textbf { B }& \textbf{C} & \textbf {D}  & \textbf{ A}&  \textbf {E}\\ 
\hline
 $A_u$(1) & 139  &  0 &   2.8 &   0  &  \color{gray}139  &  143  &  \color{gray} 139    &  141 &  143   &   \color{gray}139 & 141\\ 
 $A_u$(2) & 174  & 0  &  9.9  & 0   &  \color{gray} 174 &  \textbf{239}  &  \color{gray} 174    & 171  &  \textbf{226}   &  \color{gray} 174 & 171\\
 $A_u$(3)& 221  & 0  & -2.2  &  0   &  \color{gray} 221  &  217  & \color{gray}221      & \color{gray}221  &  211   &  \color{gray}221  &\color{gray} 221\\
 $A_u$(4)& 255  &  0 &  -2.6  &  0   &  \color{gray}255  &  267  &  \color{gray}255    &  252 &  261   &  \color{gray} 255 & 245\\
 $A_u$(5)& 284  &  0 &  -4.6  &   0  & \color{gray} 284  & \textbf{412 }  & \color{gray}284     &  274 &   306  &  \color{gray} 284  & 303\\ 
 $A_u$(6)& 302  & 0  &  -0.3  &   0 &    \color{gray}302 & \color{gray} 302  &   \color{gray} 302   & \color{gray} 302 & \color{gray} 302   &  \color{gray}302  & \color{gray}302\\
 $A_u$(7) & 339  &  0 &  -3.6  & 0   &  \color{gray}339  & 317   &  \color{gray} 339   & 329  &  347   &  \color{gray} 339  &342 \\
 $A_u$(8)&365   & 0  & -2.3   & 0   &  \color{gray}365  &  358  & \color{gray} 365    & \color{gray}365  &\color{gray} 365    &  \color{gray} 365  &\color{gray} 365\\
 $A_u$(9)& 659  & 0  & 11.8   &   0  &   \color{gray}659 & \textbf{ 941}  & \color{gray}659     & 650  &  \textbf{962}   &  \color{gray}659  &\textbf{956} \\ 
 $A_u$(10)&  737 &  0 & -2.1   &   0  &   \color{gray}737  &  734  &  \color{gray}  737   &\color{gray} 737  & \color{gray}737    & \color{gray} 737  &725 \\
  $A_u$(11) &  1020 & 0  &  0.4  &  0   &   \color{gray}1020 &\color{gray} 1020   &   \color{gray} 1020  &\color{gray} 1020 &\color{gray} 1020    & \color{gray} 1020  &\color{gray}1020 \\
  \cline{1-12}
 $B_u$(1)& 166  & 12.1  &   0 & 0.0  & \textbf{413}  &  \color{gray}166  & \color{gray} 166   & \textbf{412} & \color{gray}166   & \textbf{289}  & \textbf{262}\\ 
 $B_u$(2)& 241  & 3.9  &   0 & -2.9  & 232  &  \color{gray}241  & 248    & 234 &  246  & \color{gray}241  & \color{gray}241\\  
 $B_u$(3)& 272  &  -0.1 &   0 & 3.1  &\color{gray} 272  &  \color{gray}272   &   290  & \color{gray}272 & 277  & 251  & 277\\ 
 $B_u$(4)& 318  & 2.9  &   0 & -1.5  &  301 &  \color{gray} 318  &   321  & 308 & 321   &\color{gray}318   &\color{gray}318 \\ 
 $B_u$(5)& 345  & -0.8  &   0 &  -1.0 &  341 &  \color{gray} 345  & \color{gray}345    & \color{gray}345 &  \color{gray}345  &\color{gray}  345 &\color{gray} 345\\ 
 $B_u$(6)& 360  &  3.3 &   0 & 5.1  &349   &  \color{gray}360  &   \textbf{410}  & 353 &  \textbf{411}  & \textbf{426}  &\textbf{421} \\ 
 $B_u$(7)& 453  & -0.4  &   0 &  0.3 & \color{gray}453  &  \color{gray}453  & \color{gray} 453   & \color{gray} 453& \color{gray} 453  & \color{gray}453  & \color{gray}453\\ 
 $B_u$(8)& 640  & 12.5  &   0 & -0.1  &\textbf {952}  & \color{gray} 640  & \color{gray}  640  & \textbf{947} & \color{gray}640   &  \textbf{962} & 649\\ 
 $B_u$(9)& 766  & 0.3  &   0 & -11.1  &\color{gray} 766  & \color{gray} 766  &   \textbf{978}  &\color{gray} 766 &  709  & 706  & 739\\ 
 $B_u$(10)& 1041  & -0.2  &   0 &  -0.5 &\color{gray} 1041 & \color{gray}1041  &  1043   & \color{gray}1041 &  \color{gray}1041  & 1043  &\color{gray}1041 \\ 
       
\end{tabular}
\end{ruledtabular}
\end{table*}
\begin{table*} [h]
\caption{\label{tab:s5} The LO frequencies (in cm$^{-1}$) of the infrared active modes along the different high-symmetry directions of the Brillouin zone, their corresponding irreducible representations, and mode effective charges ($\overline{Z}^*$) of the TO infrared active modes along the $x$, $y$ and $z$ is presented for the $P2_1/n$ phase calculated by LDA. The frequencies shown by light gray in the LO columns are equal to the TO frequencies. The frequencies shown in bold font present the strong shifts in LO-TO splitting. The coordinates of the high-symmetry points are as follows: Y(1/2,0,0), Z(0,1/2,0), B(0,0,1/2), C(1/2,1/2,0), D(0,1/2,1/2), A(1/2,0,1/2) and E(1/2,1/2,1/2).}
\begin{ruledtabular}
\begin{tabular}{cccccccccccc}
\textbf{Irrep} &\textbf{TO Modes} & \multicolumn{3}{c}{\textbf{$\overline{Z}^*$}}  & \multicolumn{7}{c}{\textbf{LO Modes }} \\
\cline{3-5} \cline{6-12}
& &\textbf {x} &\textbf{y} &\textbf{z} &\textbf{Y} & \textbf {Z}  &\textbf { B }& \textbf{C} & \textbf {D}  & \textbf{ A}&  \textbf {E}\\ 
\hline
$A_u$(1) & 46 &  0 &  0.2
 &   0  &   \color{gray}46  &\color{gray} 46
  &   \color{gray}46   &\color{gray} 46
  & \color{gray}46
    &  \color{gray}46  &\color{gray} 46
\\ 
$A_u$(2) & 57  &  0 & -0.3
   &   0  &  \color{gray} 57 & \color{gray} 57
 &   \color{gray} 57   & \color{gray} 57
 & \color{gray}57
    &   \color{gray} 57 &\color{gray} 57
\\ 
$A_u$(3) & 74  &  0 & -0.6
   &   0  &   \color{gray}74  & \color{gray} 74
 &  \color{gray}74   &\color{gray}74
   & \color{gray} 74
   &  \color{gray}74   &\color{gray}74
 \\ 
$A_u$(4) &  108 &  0 & -0.04
   &   0  &  \color{gray}108  & \color{gray} 108
 &  \color{gray}108   & \color{gray}108
  &\color{gray} 108
    & \color{gray}108   & \color{gray}108
\\ 
$A_u$(5) & 134  &  0 & -0.09
   &   0  &  \color{gray}134  &\color{gray}134
   &  \color{gray}134    & \color{gray}134
  &  \color{gray} 134
  &   \color{gray}134 & \color{gray}134
\\ 
$A_u$(6) & 206 &  0 & -0.3
   &   0  &  \color{gray} 206 & \color{gray} 206
 & \color{gray} 206     & \color{gray}206
  &\color{gray}206
     &   \color{gray} 206  &212
 \\ 
$A_u$(7) & 226  &  0 &  -6.9
  &   0  &  \color{gray}  226  & 236
  & \color{gray}  226    &215
   &   233
  &  \color{gray}  226 &\textbf{397}
 \\ 
$A_u$(8) & 257  &  0 &  7.2
  &   0  &  \color{gray} 257  &\textbf{ 302}
  & \color{gray} 257    & 246
  &  264
   &  \color{gray} 257  & 242
\\ 
$A_u$(9) & 266  &  0 & 1.3
   &   0  &  \color{gray} 266& \color{gray} 266
 & \color{gray} 266    &\color{gray} 266
  &  269
   &   \color{gray} 266 &\color{gray} 266
\\ 
$A_u$(10) & 272  &  0 &  -2.8
  &   0  &  \color{gray}272  & \color{gray}272
  & \color{gray}272     & \color{gray}272
  & 275
    &  \color{gray}272  &269
 \\ 
$A_u$(11) & 308  &  0 &  1.3
  &   0  &  \color{gray}308   & 314
  & \color{gray}308    &\color{gray} 308
  & \color{gray} 308
   &  \color{gray}308   &\color{gray}308
 \\ 
$A_u$(12) & 326  &  0 &  0.4
  &   0  & \color{gray}326   & \color{gray} 326
 & \color{gray}326    &\color{gray}  326
 &\color{gray}  326
   & \color{gray}326   &\color{gray}326
 \\ 
$A_u$(13) & 333  &  0 & 1.4
   &   0  & \color{gray} 333  & 335
  & \color{gray} 333     & 329
  & \color{gray} 333
   & \color{gray} 333  & \color{gray}333
\\ 
$A_u$(14) & 346  &  0 &  -2.2
  &   0  &  \color{gray} 346   & 349
  &\color{gray} 346      & 350
  & 351
    & \color{gray} 346  & 343
\\ 
$A_u$(15) & 362  &  0 &  7.3
  &   0  & \color{gray}   362 & \textbf{410}
  &  \color{gray}   362    & \textbf{407}
  &  397
   &   \color{gray}   362 & 352
\\ 
$A_u$(16) & 425  &  0 &  0.03
  &   0  & \color{gray}425    & \color{gray}425
  & \color{gray}425     & \color{gray}425
  & \color{gray} 425
   &  \color{gray}425   & \color{gray}425
\\ 
$A_u$(17) &  446 &  0 &  0.06
  &   0  &  \color{gray}446  & \color{gray}446
  & \color{gray}446     & \color{gray}446
  &  \color{gray} 446
  &  \color{gray}446   & \color{gray}446
\\ 
$A_u$(18) &609   &  0 & 0.3
   &   0  & \color{gray} 609   & \color{gray} 609
 &  \color{gray} 609    & \color{gray}609
  &  \color{gray} 609
  &  \color{gray} 609 &\color{gray} 609
\\ 
$A_u$(19) & 695  &  0 & -13.9
   &   0  &  \color{gray} 695 &\textbf{ 957}
  & \color{gray} 695    &\textbf{958}
   &  698
   & \color{gray} 695  &\textbf{963}
 \\ 
$A_u$(20) &  699 &  0 & -7.0
   &   0  & \color{gray}699    &\color{gray} 699
  & \color{gray}699     & \color{gray}699
  &  722
   &\color{gray}699   &699
 \\ 
$A_u$(21) & 766  &  0 &  6.3
  &   0  & \color{gray}  766   & 754
  & \color{gray}  766      &756
   & \color{gray} 766
   &  \color{gray}  766  &\color{gray} 766
\\ 
$A_u$(22) & 1010  &  0 & 0.8
   &   0  &  \color{gray}1010  & 1013
  & \color{gray}1010     &1012
   &   1012
  &  \color{gray}1010 & \color{gray}1010
\\ 
$A_u$(23) & 1015 &  0 & -0.01
   &   0  &  \color{gray}1015  &\color{gray} 1015
  & \color{gray}1015    &\color{gray}1015
   & \color{gray} 1015
   & \color{gray}1015   & \color{gray}1015
\\ 
  \cline{1-12}
 $B_u$(1) & 126  & 19.5
  &   0 &  -0.06
 & \textbf{401}  &  \color{gray}126  & \color{gray}126
    &\textbf{337}
  &\color{gray} 126
   & \textbf{390}
 &\textbf{183}
\\ 
 $B_u$(2) &  194 & 0.1
  &   0 & -0.4
  & \color{gray}194
  & \color{gray} 194 &\color{gray}194
     & \color{gray}194
 & \color{gray} 194
  &\color{gray} 194
 &\color{gray}194
\\ 
 $B_u$(3) & 203  & 8.3
  &   0 &-0.5
   &  193
 & \color{gray}203  &\color{gray}203
     & 186
 & \color{gray} 203
  &191
  &206
\\ 
 $B_u$(4) & 221  &  -2.0
 &   0 & -1.2
  &\color{gray} 221
  &\color{gray} 221   &\color{gray} 221
    &\color{gray} 221
 &  \color{gray} 221
 & 218
 &\color{gray}221
\\ 
 $B_u$(5) &235   &  5.5
 &   0 & -1.1
  &  231
 & \color{gray} 235 &\color{gray} 235
    & 232
 & \color{gray}235
   & 231
 &233
\\ 
 $B_u$(6) &  265 & -0.7
  &   0 & 1.4
  &\color{gray} 265
  &\color{gray} 265    & \color{gray} 265
   &\color{gray} 265
 & \color{gray} 265
  & \color{gray}265
 &\color{gray}265
\\ 
 $B_u$(7) & 280 &  2.8
 &   0 & -2.3
  &  275
 &\color{gray}280     &\color{gray} 280
    & 278
 & \color{gray} 280
  &\color{gray} 280
 &\color{gray}280
\\ 
 $B_u$(8) & 283  & -1.8
  &   0 & -4.1
  &\color{gray} 283
  & \color{gray}283   & 291
    &\color{gray}283
  &   293
 & 257
 &275
\\ 
 $B_u$(9) & 303  &0.01
   &   0 & 0.04
  & \color{gray} 303& \color{gray} 303  &\color{gray} 303
    & \color{gray}303
 & \color{gray}303
   & 297
 &\color{gray}303
\\ 
 $B_u$(10) & 310  &  0.6
 &   0 & 5.0
  &\color{gray} 310
  & \color{gray} 310  &  314
   &\color{gray} 310
 & 314
   & 303
 &300
\\ 
 $B_u$(11) &315   & 1.8
  &   0 &  -1.1
 & \color{gray} 315
 &\color{gray}  315  &  322
   &\color{gray}315
  &  320
  & \color{gray} 315
&\color{gray}315
\\ 
 $B_u$(12) &332   &  0.3
 &   0 & 4.4
  & \color{gray} 332
 & \color{gray}332   &\textbf{384}
     &\color{gray} 332
 &   341
 &324
  &324
\\ 
 $B_u$(13) & 344  & -0.7
  &   0 & 0.8
  &\color{gray} 344
  &\color{gray} 344   & \color{gray} 344
   & \color{gray}344
 &\color{gray} 344
   & \color{gray}344
 &\color{gray}344
\\ 
 $B_u$(14) & 362  & -1.5
  &   0 & -0.3
  & \color{gray}362
   & \color{gray} 362  &\color{gray} 362
    & \color{gray}362
 &  \color{gray} 362
 &\color{gray} 362
 &\color{gray}362
\\ 
 $B_u$(15) &379   &  0.2
 &   0 & 0.6
  & \color{gray} 379
  & \color{gray}379   &376
     &\color{gray}379
  &  \color{gray} 379
 &\color{gray} 379
 &\color{gray}379
\\ 
 $B_u$(16) &  407&  -0.04
 &   0 & 1.8
  & \color{gray}407
  & \color{gray}  407 &413
     &\color{gray} 407
 &   413
 & 411
 &411
\\ 
 $B_u$(17) & 625  & 18.1
  &   0 & 0.6
 & \textbf{959}
 &\color{gray}625   &\color{gray} 625
    & 664
 & \color{gray} 625
  &\textbf {700}
 &660
\\ 
 $B_u$(18) & 756  & 0.6
  &   0 & -14.6
  &\color{gray}756
   & \color{gray}756 &\textbf {975}
    & 758
 & \textbf {967}
  & \textbf{966}
 &729
\\ 
 $B_u$(19) & 776  & 0.3
  &   0 &  -6.5
 &\color{gray} 776
  & \color{gray} 776 & 773
    &\color{gray} 776
 & \color{gray} 776
  & 773
 &\color{gray}776
\\ 
 $B_u$(20) & 847  & 0.5
  &   0 & 0.4
  & \color{gray} 847
 & \color{gray} 847  &\color{gray}847&\color{gray} 847 & \color{gray}847
   & \color{gray}847
 &\color{gray}847
\\ 
 $B_u$(21) &1042   & 0.02
  &   0 & 0.06
  & \color{gray} 1042
 & \color{gray} 1042  &\color{gray}1042 &\color{gray}1042
  & \color{gray} 1042
  &\color{gray} 1042
 &\color{gray}1042
\\ 
 $B_u$(22) & 1048 & -0.03
  &   0 & -0.9
  & \color{gray} 1048
 & \color{gray} 1048 & 1050
    & \color{gray}1048
 &  1050
  & 1050
 &1050
\\ 
\end{tabular}
\end{ruledtabular}
\end{table*}

\clearpage

\bibliography{main.bib}

\end{document}